\documentclass[journal,twoside,web]{ieeecolor}
\usepackage{generic}
\usepackage{cite}
\usepackage{amsmath,amssymb,amsfonts}
\usepackage{graphicx}
\usepackage{textcomp}
\usepackage{epstopdf}
\usepackage{color}
\usepackage[noend]{algpseudocode}
\usepackage[ruled]{algorithm2e}

\usepackage{comment}
\usepackage{epstopdf}
\usepackage[colorlinks,
linkcolor=blue,
anchorcolor=blue,
citecolor=blue]{hyperref}
\usepackage[tight,footnotesize]{subfigure}
\usepackage{amsfonts,amssymb} 
\usepackage{ulem}

\def\BibTeX{{\rm B\kern-.05em{\sc i\kern-.025em b}\kern-.08em
    T\kern-.1667em\lower.7ex\hbox{E}\kern-.125emX}}
\begin{document}
\title{Blockchain-Based Federated Learning in Mobile Edge Networks with Application in Internet of Vehicles}
\author{Rui~Wang, \IEEEmembership{Senior Member, IEEE}, Heju~Li, \IEEEmembership{Student Member, IEEE}, Erwu~Liu, \IEEEmembership{Senior Member, IEEE}
\thanks{Rui~Wang, Heju~Li, and Erwu~Liu are with 
the School of Electronics and Information, Tongji University, Shanghai 200092, China(e-mail: ruiwang@tongji.edu.cn; 2010456@tongji.edu.cn; erwuliu@tongji.edu.cn).}}

\maketitle

\begin{abstract}
The rapid increase of the data scale in Internet of Vehicles (IoV) system paradigm, hews out new possibilities in boosting the service quality for the emerging applications through data sharing. Nevertheless, privacy concerns are major bottlenecks for data providers to share private data in traditional IoV networks. To this end, federated learning (FL) as an emerging learning paradigm, where data providers only send local model updates trained on their local raw data rather than upload any raw data, has been recently proposed to build a privacy-preserving data sharing models. Unfortunately, by analyzing on the differences of uploaded local model updates from data providers, private information can still be divulged, and performance of the system cannot be guaranteed when partial federated nodes executes malicious behavior. Additionally, traditional cloud-based FL poses challenges to the communication overhead with the rapid increase of terminal equipment in IoV system. All these issues inspire us to propose an autonomous blockchain empowered privacy-preserving FL framework in this paper, where the mobile edge computing (MEC) technology was naturally integrated in IoV system. Specifically, we introduce differential privacy techniques to prevent privacy concerns while the honesty of participants are ensured by a malicious updates remove algorithm based on self-reliability filter. Simultaneously, a double aggregation frame is proposed, to guarantee the communication overhead and ensure the quality of model training. We propose to build a federated learning participant management system by employing a blockchain architecture, and ensure the immutability of uploaded model via computing and recording their quality in blockchain. The model quality is quantified as aggregate weight and used as a criterion for the distribution of federal profits, thus attracting vehicles with high-quality data to join federated learning. Numerical results derived from real-world datasets demonstrate that the proposed FL scheme for IoV system achieves good accuracy, high robustness, and enhanced security.
\end{abstract}


\section{Introduction}
\label{sec:introduction}
\IEEEPARstart{N}{ext}-generation Internet of Vehicles (IoV) system where the amount of data generated by the connected vehicles grows at a terrifying rate, can guarantee ultra-high reliable connectivity as well as low-delay response anytime, anywhere and on-the-move \cite{feng2019toward}, \cite{sun2019blockchain} foreseeingly through data sharing. Traditional cloud-centric sharing approachs request data collected from connected vehicles, e.g., measurements, videos, photos, etc., to be uploaded and processed centrally by a cloud-based server within which the processed data are futher aggregated. It can provide insights or produce effective inference models for solving the actual problems potentially emerging in IoV system, e.g., driver state analysis, road condition analysis, etc. Nevertheless, along with the benefits the data brings, some serious issues comes when demands data sharing, i.e., trading data and exchangeing knowledge between vehicles. Firstly, privacy concerns arising within the storage, transmission and sharing of private data, may touch off serious risks for data providers due to the leakage of sensitive information. 
Secondly, traditional cloud-centric sharing approach involving long propagation delays thus incuring unacceptable latency \cite{2017A}, is unfriendly with some applications in which real-time decisions have to be made, e.g., autonomous vehicles. Additionally, transferring data to the cloud not only burdens the backbone networks, but also leads to large computation and storage overhead, especially in tasks involving unstructured data, e.g., in video analytics \cite{2018Learning}. 
Such malfunction mostly brings an inaccurate model, increases the response delay and engenders distortion over model training. Therefore, it is essential for the IoV system to solve these serious concerns, via designing an efficient data-sharing mechanism.
\begin{figure*}[t]
	\centerline{\includegraphics[width=0.95\linewidth]{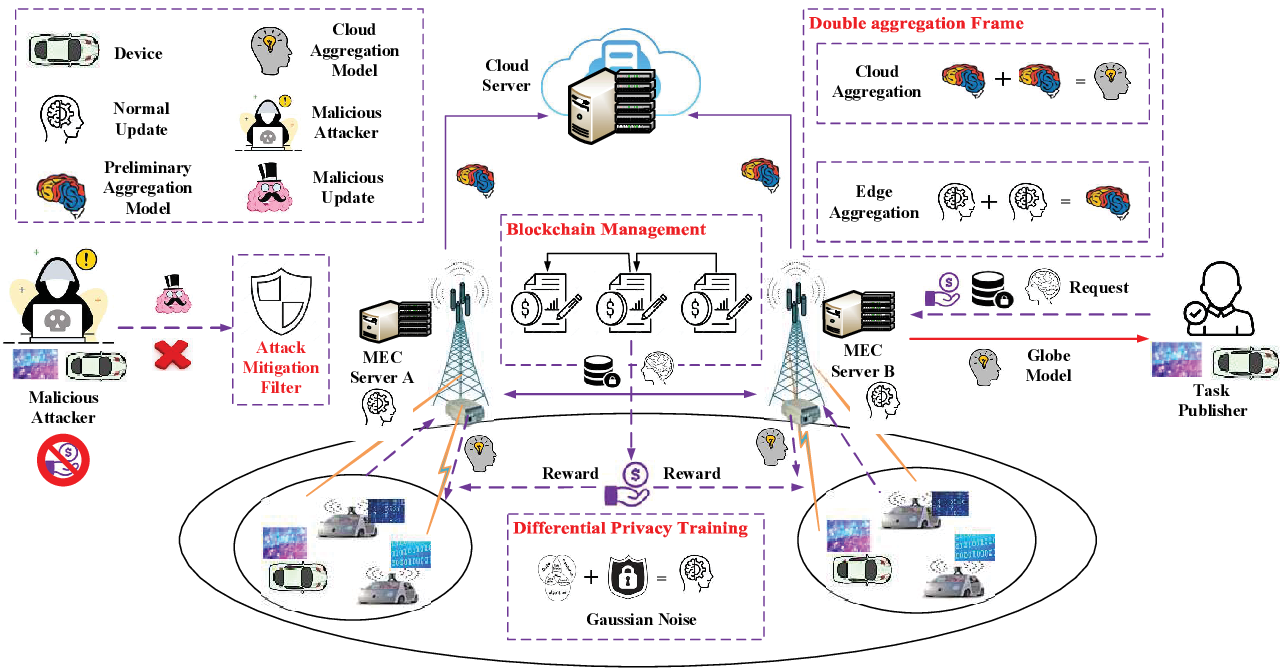}}
	\caption{The working mechanism of proposed BMFL, A blockchain-based FL model for IoV system, within which the MEC was naturally integrated.}
	\label{fig:system}
\end{figure*}

Recently, the concept of federated learning (FL), as a promising approach to address the constraints in privacy, delay and computing or storage resource, by limiting the  data scale transferred and accelerating learning processes within vehicles, has been proposed by google \cite{2016Federated} and drawn increasing attention. Rather than upload and process private data in a traditional centralized way, vehicles can simply send local model updates trained using local private data to a data center without uploading any raw data, thus decoupling the machine learning from acquiring, storing and training data in a central server. Massive decentralized data collected by vehicles can be trained individually and complementary knowledge can be accomplished among distributed model trainers. However, for designing a efficient and accurate IoV system, traditional FL is can not be applied straightforwardly due to following fundamental shortcomings \cite{9079513}:
\begin{itemize}
\item \textbf{\textit{Honesty}}: Most existing works generally suppose privacy inferring from curious participants in FL, but neglect the reality where other security threats from dishonest behaviors may exist during gradient collecting, thus disrupting the FL training process. For instance, dishonest participants may replace the updating model with its exquisitely designed attack model to poison the collaborative model \cite{bagdasaryan2020backdoor}, thus either preventing the global model from converging or leading to a sub-optimal minimum. Most stochastic gradient descent (SGD) based FL algorithms where the centralized aggregator averages the local updates to obtain the global update, are vulnerable to these malicious threats even attackers account for only a small proportion. The lack of protection severely injures the confidence of FL participants and prevents FL system from being used in many sensitive applications, e.g., home monitoring and self-driving cars. 
\item \textbf{\textit{Communication overhead}}: Most existing FL frames treat each connected vehicle as mobile data centers and allow vehicles to jointly train a shared global model in a decentralized manner, thus decoupling the machine learning from acquiring, storing and training data in a central server and exchanging for relatively lower response delay and communication burden. However, existing cloud-based aggregation approach is so sluggish that communication overhead issue may be inevitably arised with the rapid increase of terminal vehicles in IoV system. Even though the latency can be guaranteed through sinking the cloud to the edge node \cite{2019Client}, the finite number of vehicles each edge server can access, leading to the loss of training performance unavoidably. 
\item \textbf{\textit{Incentive mechanisms}}: Most existing works assume optimistically that all the vehicles in IoV system will participate in FL tasks unconditionally when were invited. However, it is not practical in the real world due to resource costs emerged during model training. For example, vehicles with high-quality data may be reluctant to participate in FL training with other vehicles with low-quality samples,  because they get the same benefits from FL tasks if without a incentive mechanism that maps the contributed resources into appropriate rewards. 
\end{itemize}

To address above issues, we propose an autonomous blockchain empowered privacy-preserving FL framework, named BMFL, where the mobile edge computing (MEC) technology was naturally integrated, as illustrated in Figure~\ref{fig:system}. More specifically, we introduce differential privacy (DP) techniques to design a privacy-preserving FL framework while the honesty of participants are ensured by a malicious updates remove algorithm based on self-reliability filter to effectively avoid the influence of false updates provided by dishonest participants. In addition, a blockchain architecture is employed to store the information of each participants, select federated learning trainers, and ensure the transparency and traceability of FL process and final profit distribution. Simultaneously, MEC technology is introduced to sink the aggregation center from cloud to the MEC server, to minimize the cloud communication overhead as much as possible, while the double aggregation frame is proposed to ensure the quality of model training. To summarize, the key contributions of this paper are shown as follows: 
\begin{itemize}
	\item To the best of our knowledge, we are the first to integrate the blockchain framework and MEC technology into privacy-preserving FL process to ensure the privacy, quality and communication overhead for the IoV system.
	\item We propose a malicious updates remove algorithm based on self-reliability filter to bias the aggregation in FL, thus effectively defensing against dishonest behavior. 
	\item Considering the contribution of each participants to the global model, we design a incentive mechanism based on model quality which is quantified as aggregate weight, to attract vehicles with high-quality data to join FL tasks. Simultaneously, we utilize the blockchain to record the model quality of each trainer model during whole training process, thus ensuring the transparency and auditability of profit distribution. 
	\item We introduce MEC technology to sink the FL aggregation center from cloud to the MEC server, to minimize the cloud communication overhead as much as possible. Simultaneously, we propose a double aggregation frame to ensure the quality of model training. 
	\item We implement BMFL prototype, and evaluate its performance with benchmark, open real-world datasets for data categorization.
\end{itemize}

The rest of this paper is organized as follows. We present the underlying concepts used in BMFL in Section II, followed by the summary of proposed FL frame in Section III. A privacy-preserving, robust and low-overhead FL framework for IoV system is described in detail in Section IV. Then the performance analysis for our proposed scheme is provided and illustrative numerical results is presented in Section V. Finally, we provide a conclusion of our work in Section VI.

\section{Background}
This section provides brief discussions on the underlying concepts which form the building blocks of the proposed BMFL framework, including the basic principles of federated learning, blockchain framework and mobile edge computing technology.
\subsection{Brief Introduction to Federated Learning}
Serving as the cross-device distributed learning system  which harnesses the benefits of low-latency, low power consumption as well as preserving users' privacy, federated learning learns a shared global model while no data ever leaves each device by coordinating massive distributed devices. Considering $m=1,2,...,M$ indicate the index of participants in FL-based IoV system and participant holds a set of private data $O_m$, the local loss function of the model vector $w$ on $O_m$ can be expressed as 
\begin{equation} 
F_{m}({w}) = \frac {1}{|{\mathcal{O}}_{m}|} \sum _{({x}_{j}, y_{j}) \in {\mathcal{O}}_{m}} f(({w}, {x}_{j}), y_{j}).\label{1}
\end{equation}
where $f(({w}, {x}_{j}), y_{j})$ represents a sample-wise loss function, which quantifies the prediction error of model $w$ on the training sample $x_j$, w.r.t. its actual label $y_j$. The complex learning problem is usually solved by SGD. Denote $iter$ as the index for the update step, and $\eta$ as the size of gradient descent, i.e., learning rate, for moving in the direction of the opposite gradient, then the model parameters are updated as
\begin{equation} 
w_m^{iter} = w_m^{iter-1} - \eta \nabla w_m^{iter - 1}. \label{SGD}
\end{equation}
where $\nabla$ indicates gradient operation. All participants are connected via a aggregator, and seek to search the model parameter $w$ collectively that minimizes the empirical risk
\begin{equation} 
\mathop {\min }\limits_w \{ F(w) \buildrel \Delta \over = \sum\limits_{m=1}^M {{p_m}{F_m}(w)} \}.\label{agg}
\end{equation}
where $p_m$ is the weight of the $m$-th participant where ${p_m} \ge 0$ and $\sum\limits_{m = 1}^M {{p_m}}  = 1$. 

For the federated learning on distributed IoV system, network bandwidth turns into the main bottleneck to globally aggregate the local updates uploaded by each participants. Federated averaging (FedAvg), as a communication-efficient iterative model averaging approach, is thus proposeed in \cite{2016Communication} to reduce the communication rounds between participants and the aggregator, as well as address the not independent and identically distributed (non-IID) data distributions by directly averaging the locally computed updates after every $E$ steps of gradient descent on each participant. Specifically, at the ${{iter}}$ iteration, the aggregator selects a subset of participants $i \in {\Im ^{iter}} \subseteq \{ 1,2,...,M\}$ in the proportion of $V$, where $|{\Im ^{iter}}| = M * V = P$. If $iter|E \ne 0$ where $|$ is a divisional symbol, each selected vehicle $i \in {\Im ^{iter}}$ runs stochastic gradient algorithm algorithm based on its local dataset ${O}_{m}$ and the global model ${w}^{iter-1}$ transferred by the aggregator in iteration where $iter|E = 0$, and output the local updates ${w}_i^{iter}$ using Equation~(\ref{SGD}). If and only if $iter|E = 0$, the local model ${w}_i^{iter}$ will be upload to the aggregator by each selected participant. Conclusively, the aggregation center aggregates all the local updates ${w}_i^{iter}$ with $i \in {\Im ^{iter}}$, and evolves the updated local model ${w}_i^{iter}$ by
\begin{equation} 
w^{iter} = \frac{1}{P}\sum\limits_{i = 1}^{P} {[w_i^{iter - 1} - {\eta}\nabla w_i^{iter - 1}]}.
\end{equation}

\subsection{Brief Introduction to Blockchain Technology}
Blockchain is an emerging technology as a decentralized, immutable, sharing and time-ordered ledger, which is maintained by all the nodes in a peer-to-peer network, i.e. blockchain network. Typically, blockchains can be classified into two categories: permissionless blockchain and permissioned blockchain. With permissionless blockchains, such as Bitcoin \cite{nakamoto2019bitcoin} and Ethereum \cite{wood2014ethereum}, participants are allowed to join and leave at anytime and the number of participants is not pre-defined nor fixed \cite{9098045}. Permissioned blockchains (a.k.a. consortium blockchains), e.g., IBM's Hyperledger Fabric \cite{androulaki2018hyperledger}, allows only the authorized entities to join and leave the network, participate in the consensus process, send transactions, and maintain the shared ledger. 

Transactions that contain timestamps and references (i.e., the hash of previous block) are stored in blocks,  resulting an ordered list of blocks. For instance, transactions in Bitcoin referring to money transfers, are created by pseudonymous participants and stored to build a new block competitively by an entity called consensus node. A consensus mechanism, such as proof of work (PoW)-based, proof of stake (PoS)-based \cite{larimer2013transactions}, byzantine fault tolerance (BFT)-based \cite{castro1999practical}, is executed by consensus node to compete for the power to write self-organized blocks into the blockchain. 

Blockchain is well known for its transparency, temocracy, security and immutability, i.e., data and all operations are recorded on the blockchain irreversibly in an append-only manner and are accessible by all the blockchain nodes. Intuitively, the incremental characteristic of FL in IoV system makes it suitable for leveraging blockchain to ensure data integrity, encourage vehicles to share data and enable transparent training management. However, it is quite essential to design a reasonable approach for integrating blockchain with privacy-preserving federated learning,.
\subsection{Brief Introduction to Mobile Edge Computing}
Cloud computing \cite{FERNANDO201384}, as a successful paradigm, which opens up richer and more complex applications to users with the help of the power of the  cloud server. Nevertheless, extremely sensitive latency requirements have demanded an alternate approach \cite{8016573}. Additionly, the network architecture is becoming increasingly heterogeneous due to the complex traffic distributions in wireless communications systems. To this end, MEC is naturally proposed as a novel distributed computation architecture, that deploys application servers at the edge of wireless networks closing to where data is produced, to bring powerful capabilities in real-time data storaging and processing at the edge of network. It can be seen as a footing stone that naturally bridge the cloud-based learning services and mobile devices. MEC is specially significant for accelerating the rapid download of various applications and allowing users to enjoy an uninterrupted high-quality network experience, due to the characteristics in ultra-high bandwidth, ultra-low latency, and strong real-time.

In a cloud-centric FL framework where long propagation delays and communication overhead issue may be inevitably arised with the rapid increase of terminal vehicles in IoV system, puts forward a new demand for the development of FL architecture in the IoV system. On the other hand, the limited storage and computing resources of the terminal vehicles make it insufficient as the main blockchain nodes, which need to validate all transactions generated in FL training and implement complex consensus mechanisms to competing for writing blocks. These requirements in FL framework which consistent with the definition and characteristics of MEC, naturally inspire us to integrate MEC into federated learning to obtain a low-overhead and efficient FL framework. Nevertheless, compared with cloud-centric FL framework, the limited number of clients each MEC server can access, leads to inevitable training performance loss. Therefore, the trade between communication overhead and system performance should be made appropriately.

\section{System Model}
In this section, a privacy-preserving, secure, low-overhead and motivating FL framework in IoV system, namely BMFL, is presented, where blockchain and MEC technology are integrated. Our goal is to design a secure FL framework which can ensure the privacy, robust, low communication overhead, while also can encourage vehicles in the IoV system to participate in FL process during whole FL training process.
\subsection{Related Concepts}
We first describe the related concepts introduced in the proposed BMFL framework.
\begin{itemize}
    \item \textbf{\textit{Device}}: A physical entity in IoV system involving in FL framework. In this paper, we simply treat vehicles in IoV system as a device. In blockchain architecture, a device is also called model node. 
    \item \textbf{\textit{MEC server}}: A physical entity with certain computing and storage resources in IoV system, such as base station (BS). In blockchain architecture, a MEC server is also called consensus node. 
    \item \textbf{\textit{Task publisher}}: Anyone who can publish FL tasks and interacts with the belonged MEC server.
    \item \textbf{\textit{Trainer}}: Entities whose dataset meet the task requirements and have the intention to participate in FL.
    \item \textbf{\textit{Iteration}}: Model training consists of one update step, i.e, SGD.
    \item \textbf{\textit{Aggregation interval}}: Model training consists of multiple steps, where all the weights of neurons of the model are updated once.
    \item \textbf{\textit{Local model update}}: The local trained model uploaded by a device.
    \item \textbf{\textit{Global model}}: An initial global model issued by the task publisher or global models updated by the central aggregators during an aggregation interval.
    \item \textbf{\textit{Blockchain node}}: Entities in blockchain which are motivated to generate, collect, verify, and write transactions into blocks, consisting of model nodes and consensus nodes.
    \item \textbf{\textit{Leader}}: Entity which is selected from all blockchain nodes during an aggregation interval via consensus protocol to write self-organized blocks into the blockchain.
\end{itemize}
\subsection{Threat Model}
A typical FL process, where $M$ devices (i.e., vehicles), $N$ MEC servers and one task publisher work together to accomplish a FL task, is considered in this paper. Furthermore, we define threat models in FL process as shown below:
\begin{itemize}
	\item \textbf{\textit{Threat 1: Disclosure of local data and model.}} Although each trainer only uploads local model update to the aggregator in FL framework, adversaries which can be either other trainers or MEC servers, still can infer important information about the local data \cite{2018Inference} or the sensitive information of the model \cite{hitaj2017deep} (e.g., raw distribution of training data) by initiating an inference attack. Note that all MEC servers are considered to be honest-but-curious, which means it may try to infer trainers's (not include task publisher) data privacy independently, but will execute the program according to the agreed agreement. 
	\textbf{\item \textit{Threat 2: Trainers with malicious behaviors.}}
	Malicious trainers may have intentional inappropriate behaviors during FL training, thus reducing the
	usability of the collaborative model. They may  tamper with a certain proportion of poisonous training data or wreck trainning models deliberately, to increase the probability of misclassification, thus manipulating the results of collaborative training \cite{bagdasaryan2020backdoor}.
\end{itemize}

\subsection{Our Proposed Architecture}
BMFL combines FL together with blockchain and MEC technology to achieve a privacy-preserving, secure, low-overhead and motivating data sharing framework in IoV system, in IoV system. Figure~\ref{fig:system} further illustrates the working mechanism of our proposed BMFL framework. 
In this framework, we introduce the blockchain technique to store the information of each device, select trainers automatically, and ensure the transparency and traceability of federated training process and final profit distribution. The local differential privacy technique is applied to curtail the inference attacks while the honesty of participants are ensured by a malicious updates remove algorithm based on self-reliability filter to effectively avoid the influence of false gradients provided by dishonest trainers. Simultaneously, the MEC technology is introduced to sink the aggregation center from cloud to the MEC server, to minimize the cloud communication overhead as much as possible. Furthermore, we propose an double aggregation frame, where a cloud server is deployed to re-aggregate updates generated via aggregating local model updates in massive edge servers. 

Suppose all devices have registered in a permissioned blockchain-based participant management system to turn into model nodes and obtained their unique identity (ID) by uploading blockchain node registration transactions. A task publisher which owns a small amount of testing dataset related to expected model, launches a model training request to its nearby blockchain node equipped with computing and storage resources, i.e., consensus node (MEC server), which following broadcasts this intention and testing dataset to all consensus nodes. According to the task publisher's objective, all consensus nodes select trainers towards current aggregation interval in federated training and send initial models to the trainers in charge. 

For iteration $iter$, each trainer will train the local model using its local dataset in a form of differential privacy. Once a local model is trained successfully and if and only if $iter|E = 0$, trainers submit local model updates to the nearby consensus nodes to wait for model evaluation and aggregation. Through a malicious update remove algorithm based on self-reliability filter, which aims to effectively defense against malicious attacks, each MEC server selects the most suitable updates which can potentially bring the greatest benefits to the globe model to produce preliminary aggregation model by adopting an attack-resistant aggregation model based on repeated median estimator and the reweighting scheme.  The generated preliminary aggregation models are subsequently delivered to the cloud server to re-aggregate. The re-aggregated model is used as the initial global model for the next iteration. Each model of trainers with corresponding quality will be written into the block by consensus nodes in the form of verification transactions for following audit. All transactions arising in this aggregation interval will be packaged and verified through a blockchain network, where a consensus mechanism based on the reliability of preliminary aggregation model of each consensus node in collaborative training, is executed to select a leader. Then, the leader broadcasts its self-organized block into the entire blockchain and write into the blockchain after verifying by all blockchain nodes. Once FL task is finished, the profit of this model, which is usually provided by the task publisher when initiating the model request, is partitioned to all trainers according to each trainer's contribution. In this paper, we propose a unique contribution quantification method, where the trainer's contribution is quantified as aggregate weight during training process. 

\section{Blockchain-based Federated Learning with Mobile Edge Computing}
This section details the implementation of BMFL to enhance both privacy, robustness and low-overhead as well as fair incentive. These include how to manage FL participants through blockchain, protect privacy by differential privacy algorithm, aggregate robustly through malicious updates remove algorithm and attack-resistant aggregation algorithm. We also present how to locally execute low-overhead model training by a double aggregation frame, implement consensus mechanism to maintain blockchain, followed by the quantification of fairness by incentive mechanism.
\subsection{Blockchain-based Participants Management}
\begin{figure}[!t]
	\centerline{\includegraphics[width=1\linewidth]{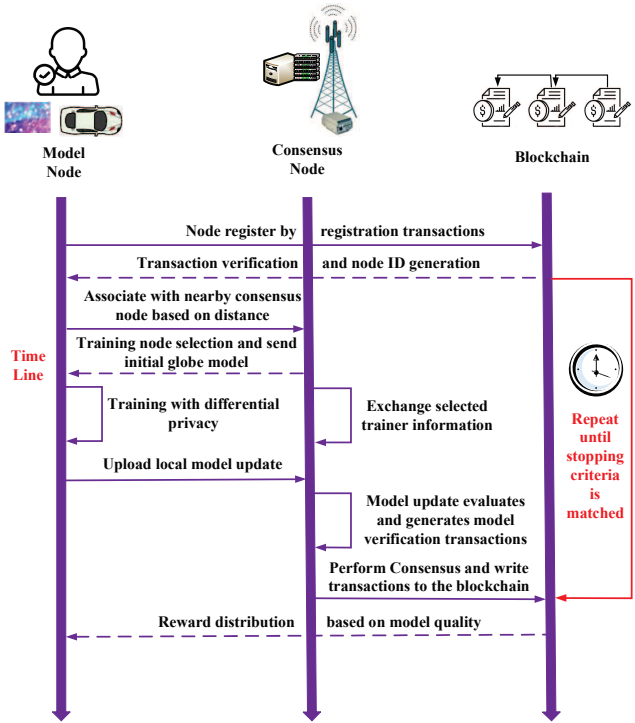}}
	\caption{Blockchain-based Architecture.}
	\label{fig:blockchain}
\end{figure}

To ensure the effective management of FL participants, thus facilitating subsequent federation training construction, incentive mechanism and responsibility traceability, we first designed a participants management system based on the permissioned blockchain as shown in Figure~\ref{fig:blockchain}, where all participants, i.e., devices and MEC servers are respectively divided into as \cite{8668426}: 
\begin{itemize}
	\item \textbf{\textit{Model node:}} Entities in blockchain which can be basically seen as devices that are responsible for providing local model update training with private data. At any point, the model nodes can be classified as two types: 1) active model nodes with model being transmitted, and 2) idle model nodes with no transmissions. Note that the two states are only logical state in blockchain and FL framework, and has nothing to do with the state of the node device in reality.
	\textbf{\item \textit{Consensus node:}} Entities in blockchain which consist of MEC servers with high computing and storage power. They have full functionalities to support blockchain protocols, and thus to take charge of transaction confirmation, data storage and generating verification transactions, as well as aggregating the verified model and implementing consensus mechanism to release block. 
\end{itemize}
\begin{figure}[!t]
	\centerline{\includegraphics[width=1\linewidth]{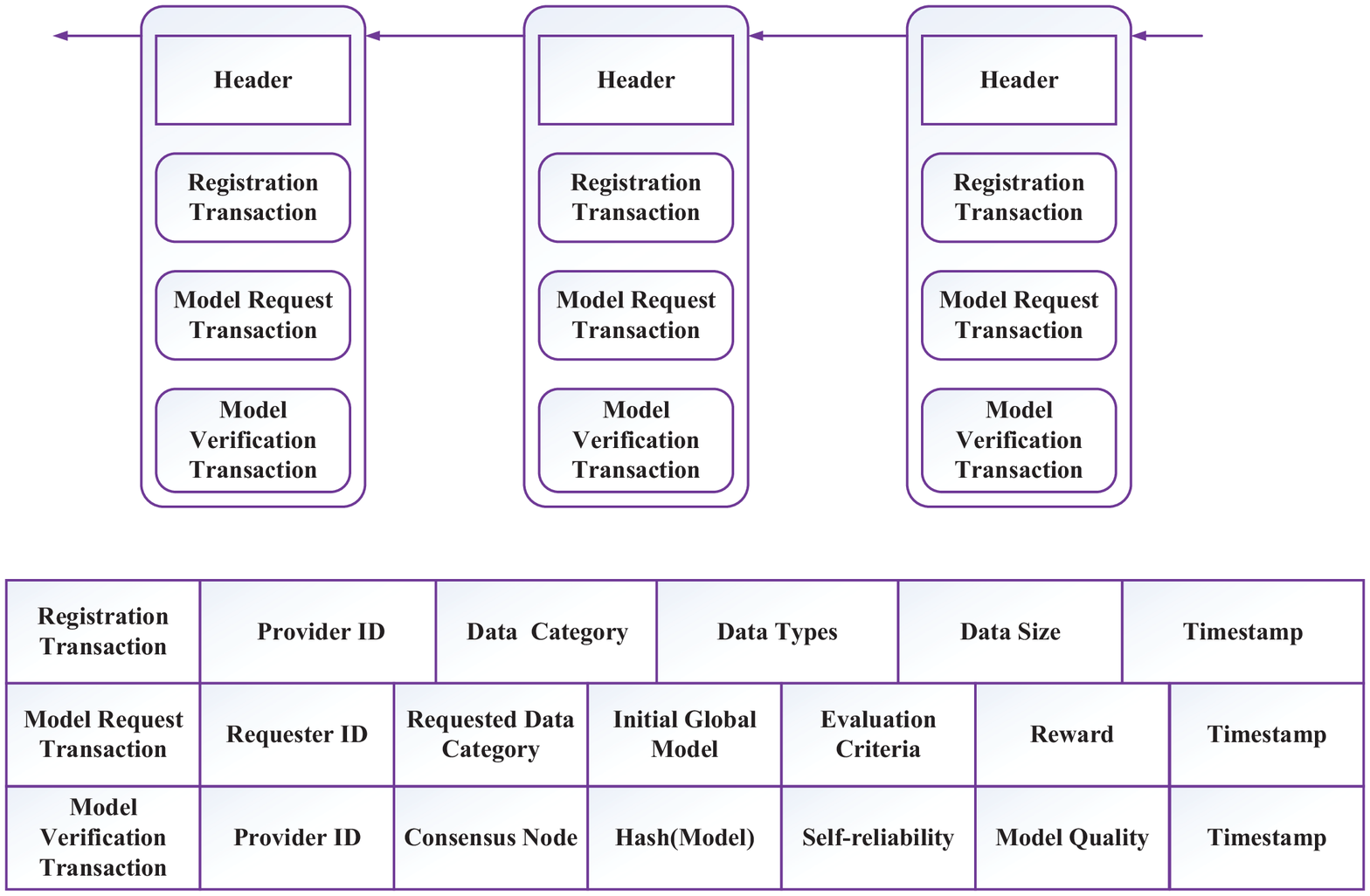}}
	\caption{The transactions of blockchain.}
	\label{fig:transaction}
\end{figure}
\begin{figure}[!t]
	\centerline{\includegraphics[width=1\linewidth]{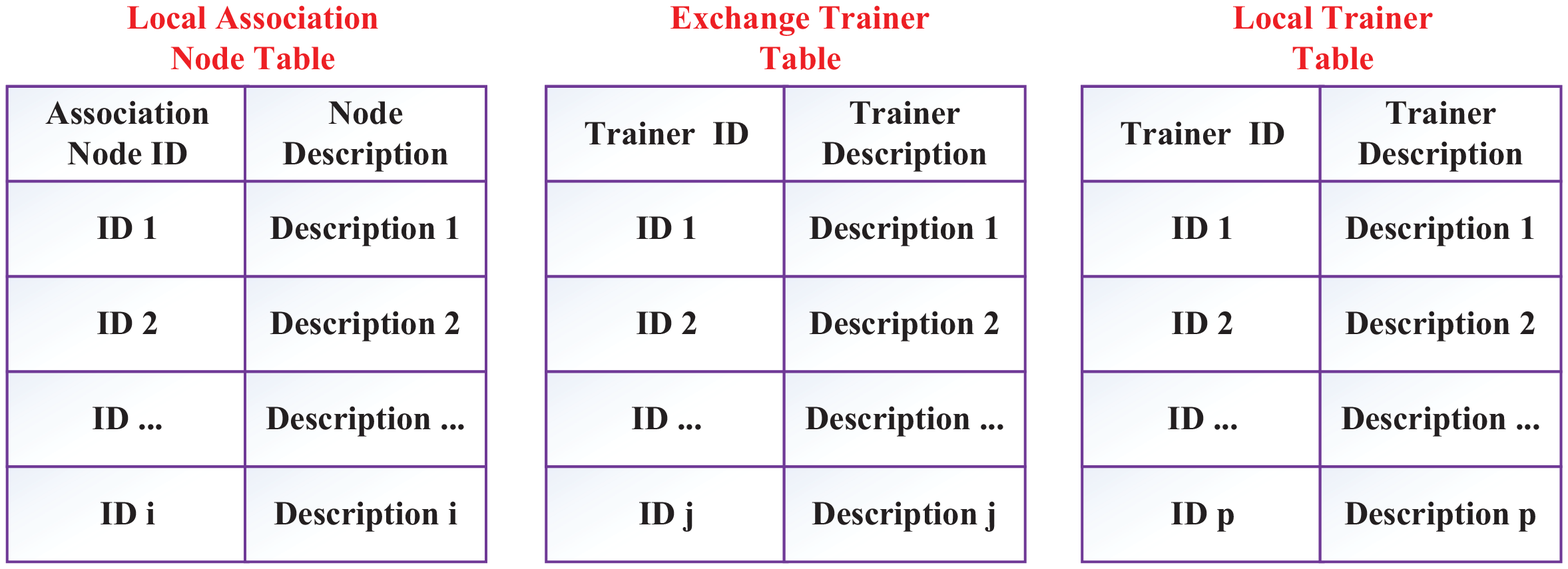}}
	\caption{The association node table and trainer information table.}
	\label{fig:MECtable}
\end{figure}

Consensus nodes are connected with each other through high data rate links via independent interface where connection types may include wired link, wireless point to point link, and wired and wireless hybrid relay link \cite{sun2019blockchain}, thus guaranteeing the security and improve the effectiveness in blockchain. The connection type between two specific consensus nodes is determined by the environment around the two nodes. For example, traditional wireless link can be adopted when the distance between nodes is short, while wired link should be used in the complicated environment. Moreover, for the large scale IoV networks, it could be more practical and cost effective to employ the wired-wireless hybrid connections. In addition, a model node can choose and associate with a suitable consensus node according to current network state through wireless communications. 

Actually, the model node set can dynamically switch depending on the needs. In other words, we utilize blockchain to provide the native ability to automatically coordinate the joining and departure of each node, further facilitating independence and modularity of the federation. When a new model node participates in, it should register with its identical information, together with the profiles of its data, including data categories, data types and data size, in forms of registration transactions in blockchain as shown in Figure~\ref{fig:transaction} and will be verified by the consensus nodes through adopting merkle tree \cite{2016Hawk}, to get its pseudonymous public key ${PK}$ and corresponding secret key ${SK}$. We adopt the identity (ID) mechanism \cite{2019Blockchain}, where each nodes holds its unique ID based on the hashed vectors, to distinguish each participating nodes (both consensus nodes and model nodes). Model nodes holding similar datasets have similar IDs to facilitate that consensus nodes can easily screen model nodes with similar datasets. 

Once registering successfully, a node turns into a model node and needs to associate with nearby consensus node. Each consensus node holds an association node table locally, which stores the main information of all model nodes associated with it, including node ID, node description, etc, as shown in Figure~\ref{fig:MECtable}. Note that the network state, including the active model nodes, the association relationship between model nodes and consensus nodes, etc. can be dynamic along with time, in other words, it changes dynamically in different FL aggregation intervals. More specifically, at the beginning of aggregation intervals, the consensus nodes, i.e., the MEC servers, will select trainers towards current training intervals in federated training according to the correlation between requirements of task requester and data owned by each model node. The selected model nodes will be activated while unselected nodes will be idle in current iteration. Generally, it is practical to assume that the minimum distance between a model node and a consensus node is set as $d_{min}$. The association rule between model nodes and consensus nodes is based on distance, i.e., a model node is associated with the nearest consensus node\footnote{In IoV system, this distance can be approximately replaced by the power of the MEC servers, i.e, the model nodes tends to associate with the consensus node with the highest instantaneous power received.}. Therefore, association relationship between model node and consensus nodes will be dynamically adjusted with the change of position of the model node, which is usually determined by the mobility of devices in the IoV system. When a model node changes its association relationship with a consensus node, it will explain the ID of the old associated consensus node to the new consensus node so that the new consensus node can notify the old node to update its association table.


A task publisher $r$ posts a FL task in forms of model request transaction shown in Figure~\ref{fig:transaction}, which is signed by $r$ with its private key ${SK}_r$, including the ID of $r$, requested data category, an initial global model including concrete collaborative model together with initial parameters, evaluation criteria, a reward amount ${\rho _{total}}$, and the timestamp, to its nearby consensus node. Generally, the accuracy performance or the maximum number of iterations are used as the evaluation criterion to control the FL training. In the presence of malicious model nodes, we assume that the task publisher has a small amount of testing dataset \cite{liu2020secure}, which will be secretly sent from the task publisher to nearby consensus node, to facilitate our subsequent malicious updates filtering (detailed in the Section IV-C). Once consensus node receives the model request, it verifies the identity of requester $r$ and broadcasts this request and testing dataset to all other consensus nodes. Note that the testing dataset should be invisible to all participating model nodes, in case the undisciplined model nodes conduct targeted training model based on the characteristics of testing dataset to confuse the subsequent evaluation. 

According to the local association table, each consensus node randomly selects partial nodes from the set of model nodes related to the request model and sends initial model parameters to them. Note that the switching of the association relationship between model nodes and consensus nodes may cause additional issue. For instance, assuming that a model node $V$ is selected by a consensus node $M_1$ to participate in FL task in a certain aggregation interval, but after the training is completed, it is already within the scope of the consensus node $M_2$, i.e., $d(V,{M_1}) \textgreater d(V,{M_2})$. At this time, if node $V$ still sends the trained model to node $M_1$, the transmission overhead may inevitably increase. Otherwise, node $V$ sends a model to $M_2$, $M_2$  may refuse to include the model in subsequent operations because it does not know if $V$ is selected as a trainer. To this end, we require each consensus node exchange the information of selected trainers with other consensus nodes, and locally generate a trainer information table in each iteration, as shown in Figure~\ref{fig:MECtable}. In this case, consensus nodes only need to check whether the received model ID exist in the trainer information table of current aggregation interval, to decide whether to include the model in subsequent evaluations and aggregations. 
\subsection{Training with Differential Privacy}
As defined in Section III-B, devices as well as the semi-honest MEC servers are trying to infer sensitive information related to local privacy of others from the uploaded model updates, by initiating an inference attack. To this end, we incorporate a differential privacy preserved mechanism into FL, thus protecting data privacy during whole decentralized learning. Differential privacy \cite{dwork2014algorithmic} trades off privacy and accuracy via perturbing the data in a way that is \cite{9098045} 1) computationally efficient, 2) does not allow an attacker to recover the original data, and 3) does not severely affect utility.

A randomized algorithm ${\rm \mathcal {A}}:\mathcal {D} \to \mathcal {R}$ with domain $\mathcal {D}$ and range $\mathcal {R}$ achieves $(\varepsilon,\delta)$-differential privacy if for any two neighbouring inputs $D$ and $D'\in \mathcal {D}$ differring in at most one record and any measurable subset of outputs outcomes $O \in \mathcal {R}$, it holds that 
\begin{equation}
Pr[\mathcal {A}(D) \in O] \leq \text{exp}(\epsilon) \cdot Pr [ \mathcal {A}(D^{\prime }) \in O]+ \delta. \label{DP} 
\end{equation}
where $\varepsilon$ refers to privacy budget, which measures the privacy leakage in differential privacy. In general, the smaller $\varepsilon$ provides higher level of privacy protection but lower accuracy. In this definition, $\delta$ accounts for the probability that the  $\varepsilon$-privacy loss exceeds. Furthermore, $\mathcal {A}$ is said to preserve $\epsilon$-differential privacy if $\delta = 0$.

The goal of DP-based federated learning is to train a global model as shown in Equation~(\ref{agg}) with an optimization problem to minimize $F(w)$ and achieve strong privacy guarantee, i.e.,
\begin{flalign}  
  \begin{split}
  &\mathop {\min}\limits_w \ F(w)\ \\
  &\mbox{s.t.}\quad\Pr ({w_m} \in {O}) \le \exp (\varepsilon)\Pr ({w_m}^\prime  \in {O}) + \delta \\
  &\;\quad\quad m \in {1,2,...,M}.\\
  \end{split}
\end{flalign} 
where $\Pr ({w_m} \in {O}) \le \exp (\varepsilon)\Pr ({w_m}^\prime  \in {O}) + \delta$ is the $(\varepsilon,\delta)$-privacy guarantee for update model $w_m$.
\begin{algorithm}[t]
	\caption{Training with Differential Privacy}
	\label{alg:DP}
	\LinesNumbered 
	\KwIn{Index of selected trainers $m = 1,2,...,M$, clipping threshold $C$, initial model ${w}^0$, loss function $F$, momentum $\beta$, noise standard deviation $\sigma \cdot C $, local batchsize $B$, number of local epochs $E$}
	\KwOut{Differential privacy model $\tilde w_m^{iter}$;}
	\For{each iteration $iter$}{
		\eIf{$iter|E = 0$}
		{Each consensus nodes send aggregate model ${w}^{iter-1}$ to the selected trainers in charge \;}
		{Each trainer compute $w^{iter-1}$ by running stochastic gradient algorithm algorithm\;}
		\For{each trainer $m$}{
			${w_m^{iter - 1}} = {w}^{iter-1}$\;
			Take a random sample $O_m^{iter} = \{ {x_1},{x_2},...,{x_B}\}$ with sampling probability
			$B/|O_m|$;\\
			\For{For each $x_i \in O_m^{iter}$}{
			Compute gradient\\
			$g_m^{iter}(x_i) = \nabla {F_m}(w_m^{iter - 1},x_i)$\;
			Clip gradient\\
			$\bar g_m^{iter}({x_i}) = \frac{{g_m^{iter}({x_i})}}{{\max (1,\frac{{{{\left\| {g_m^{iter}({x_i})} \right\|}_2}}}{C})}}$}
			Add noise\\
			$\tilde g_m^{iter} = \frac{1}{B}(\sum\nolimits_i {\bar g_m^{iter}({x_i})}  + N(0,{{\sigma ^2}{C^2}})$\;
			Add momentum\\
			$\hat g_m^{iter} = \tilde g_m^{iter} + \beta \hat g_m^{iter - 1}$\;
			Upload noised parameters as\\
			$\tilde w_m^{iter} = {w^{iter - 1}} - \eta \hat g_m^{iter}$
		}	
	}
	\textbf{Return:} $\tilde w_m^{iter}$
\end{algorithm}
For trainer $m$ in iteration $iter$, a local update model  $w_m^{iter}$ can be computed by SGD algorithm according to the equation as
\begin{equation}
w_m^{iter} = w_m^{iter-1} - \eta \hat g_m^{iter}.
\end{equation}
where $\hat g_m^{iter}$ denotes gradient of current iteration.
In this paper, $\hat g_m^{iter}$ is generated by adopting the gaussian mechanism (GM) \cite{abadi2016deep} on local gradient. More specifically, the gradient parameters, which are cliped with threshold $C$, to achieve differential privacy as described in the following equation
\begin{equation}
\bar g_m^{iter}({x_i}) = \frac{{\nabla {F_m}(w_m^{iter - 1},{x_i})}}{{\max (1,\frac{{{{\left\| {g_m^{iter}({x_i})} \right\|}_2}}}{C})}}.
\end{equation}
\begin{equation}
\hat g_m^{iter} = \underbrace{\frac{1}{B}\sum\nolimits_i {\bar g_m^{iter}({x_i})}}_{{\text{Clip gradient}}} + \underbrace {\frac{1}{B}N(0,{{\sigma ^2}{C^2}})}_{{\text{Add noise}}} + \underbrace {\beta \hat g_m^{iter - 1}}_{{\text{Add momentum}}}.
\end{equation}
where $N(0,{{\sigma ^2}{C^2}})$ indicates the added Gaussian noise with mean $0$ and standard deviation ${{\sigma}{C}}$. We set $\sigma  \geqslant \frac{{\sqrt {2\ln ({{1.25} \mathord{\left/{\vphantom {{1.25} \delta }} \right.\kern-\nulldelimiterspace} \delta })}}}{\varepsilon }$. 
Algorithm \ref{alg:DP} outlines the training algorithm with a DP guarantee.

By incorporating differential privacy in the training process, vehicle $m$ trains the noised update $\tilde w_m^{iter}$ locally in iteration $iter$, which is submitted to the nearby consensus node to compete for eligibility for model aggregation if $iter|E=0$. Therefore, adversaries can hardly recover the original model parameters and detect a slight change in the dataset by observing the output models, even if gainning access to the noise-adding gradient information, so as to achieve the purpose of protecting privacy in FL framework. In fact, another potential reason for introducing differential privacy into the training process is that DP training may limit the efficacy of a backdoor attrack by setting a low clipping bounds and high noise variance \cite{bagdasaryan2020backdoor}. Because the attack increases the distance of the backdoored model to the global model, it is more sensitive to clipping than to noise addition.
\subsection{Attack Mitigation and Double aggregation}
\subsubsection{Attacks in Federated Learning}
Federated learning which is promising to be applied to construct a collaborative model without exposing users' private data, is substantially different
from traditional centralized learning. This peculiarity determines federated learning to be utilized to coordinate thousands or millions of devices in the future IoV system to complete cooperative modeling, without restrictions on eligibility on heterogeneous data scale. Nevertheless, participants' behavior in FL training cannot be monitored unlike centralized way, leading that we can not exclude malicious participants by relying solely on the devices' autologous security guarantees \cite{bagdasaryan2020backdoor} when training with thousands of participants. The participants performing maliciously and engenderring unreliable updates may be caused by: 1) sensing data from malicious intent or tampered devices may include deceptive information, which is similar to false data injection attacks in smart grids \cite{zhuang2019false}; 2) the data can be arbitrarily manipulated when being transmitted through insecure communication channels \cite{zhu2018blockchain}. 

In FL framework, when a malicious device is selected to be a trainer, it may tamper with a certain proportion of poisonous training data or wreck trainning models deliberately, to increase the probability of misclassification, thus manipulating the results of collaborative training \cite{bagdasaryan2020backdoor}, \cite{li2019rsa}. For iteration $iter$ with $iter|E = 0$, i.e., when starting aggregation, let ${\mathbb{B}^{iter}}$ and ${\mathbb{N}^{iter}}$ indicate the set of malicious and normal trainers respectively, the update uploaded by trainer $m$, $\tilde Q_m^{iter} \in {\mathbb{R}^d}$ can be expressed as
\begin{equation}
\tilde Q_m^{iter} = \left\{ \begin{array}{l}
\tilde w_m^{iter}{\rm{\quad\;\,\, if \;}}m \in {\mathbb{N}^{iter}}\\
* {\rm{ \quad\quad\quad\; if \;}}m \in {\mathbb{B}^{iter}}
\end{array} \right.
\end{equation}
where $*$ denotes that $\tilde Q_m^{iter}$ can be an arbitrary vector in ${\mathbb{R}^d}$. 
Malicious $\tilde Q_m^{iter}$ may degrade the quality of the global model if being aggregated by a central aggregator, hence affecting the final outputs of the whole FL training. Therefore, it is vitally important to remove these malicious update before aggregation.

The universal idea of malicious updates remove strategy, e.g., 1) calculate the reliability of the model to find out abnormal prediction accuracy as proposed in \cite{liu2020secure}, where the accuracy performance on the testing dataset is used as the evaluation criterion to evaluate the reliability of models, 2)
calculate the confidence of updates based on the residual to a joint statistical properties of models after collecting the updates of all trainers, as shown in \cite{2019Attack}, where a regression line estimated by the repeated median estimator is constructed. Unfortunately, the accuracy performance on the testing dataset is effective only in the IID scenario, where the accuracy of the normal trainers is significantly better than that of the malicious trainer, as represented in \cite{liu2020secure}. 
Under non-IID environments, in the initial iteration, there is no obvious difference between the accuracy of the normal trainers and the malicious trainers, and the accuracy performance of some normal trainers is even lower than that of the malicious trainers. At this time, if the threshold-based removal method proposed in \cite{liu2020secure} is adopted, some normal trainers will be removed instead, and the malicious models of some malicious trainers will be aggregated.
The residual based reweighting aggregation algorithm \cite{2019Attack}, which calculates the confidence of updates based on the residual to a joint repeated median of all model updates, can achieve good performance in non-IID situations, however, it cannot be effectively applied to our situation where each of consensus node only retains and aggregate the upload updates of their associated trainers. Although we assume that each consensus node is responsible for a sufficient number of trainers, due to the uncertainty in the distribution of malicious trainer, e.g., the proportion of malicious trainers in the trainer-set responsible for of a certain consensus node is likely to exceed 51$\%$ (intuitively exist with a high probability due to the limitation in service area of each consensus node), the residual based reweighting aggregation algorithm used in \cite{2019Attack}, which judge the malicious degree of the participant based on the median value of all the trainer parameters, will lead that the median judgment standard inevitably fall among the parameters closing to malicious trainers. In this case, some malicious trainers will be inevitably included in the aggregation process. 
 
\subsubsection{Malicious Updates Remove Based on Self-reliability Filter}
In contrast to those methods listed above, we propose a malicious updates remove method based on the self-reliability of a specific updates, which does not depend on the model parameters of other training nodes, but is only related to the quality of its own model, thus naturally avoiding the influence of 51$\%$ malicious models proportion.

Assuming that each consensus node is responsible for a sufficient number of trainers\footnote{In fact, this assumption is very likely to be established due to the rapid increase in the volume of connected devices and data diversity in the future IoV system.}, every trainer sends its own local update (noise-adding updates for normal trainers while $*$ for malicious trainers) to nearby consensus node in each aggregation interval. When each consensus node has accumulated a certain amount of local updates, it will tend to execute malicious updates remove mechanism to eliminate the influence of malicious nodes, thereby biasing the aggregation in FL. More specifically, for iteration $iter$ where $iter|E = 0$, the consensus nodes estimates whether updates of model node $m$ is malicious or not based on the self-reliability $\Lambda _m^{iter}$, which can be calculated as 
\begin{equation}
\Lambda _m^{iter} = (1+\chi/{iter})\alpha _m^{iter} -  D_m^{iter}.
\end{equation}
where $\chi$ is a control variable related to the degree of differential privacy and local training. In this paper, we set $\chi=0.5$. $\alpha _m^{iter}$ denotes the accuracy performance of model update in testing data similar to \cite{liu2020secure}, which can be calculated as $\alpha _m^{iter} = \frac{r_t}{v}$, with $r_t$ being the number of matches between real labels of testing data including $v$ data samples, and predicted labels of model update. $D_m^{iter}$ represents the distance between the model update and the initial model (i.e., the global model in $iter-1$ iteration), which can be expressed as
\begin{equation}
D_m^{iter} = \left\{ \begin{gathered}
{\left\| {\tilde w_m^{iter} - {w^{iter - 1}}} \right\|^2}\;\;\;{\text{if}}\;\text{dir}=1 \hfill \\
\inf \quad\quad\quad\quad\quad\quad\quad\;\;{\text{if}}\;\text{dir}=-1 \hfill \\ 
\end{gathered}  \right.
\end{equation}
where $\text{dir}(-)$ is the coordinate-wise direction function, which can be calculated as
\begin{equation}
\text{dir} = \left\{ \begin{array}{l}
1{\rm{\quad\; \;\;if\; sum}}(({\rm{sign}}(\tilde w_m^{iter}(l) * w_m^{iter - 1}(l)))) > 0\\
- 1{\rm{\quad if\;sum}}(({\rm{sign}}(\tilde w_m^{iter}(l) * w_m^{iter - 1}(l)))) \le 0
\end{array} \right.
\end{equation}
where $l$ is the number of layers of model parameters and $\text{sign}()$ is the coordinate-wise sign function.
Intuitively, ${\tilde w_m^{iter}}$ is not far away the initial model ${{w^{iter - 1}}}$ to a certain extent of a normal update, although a certain degree of local training and differential privacy processing are performed locally. For a malicious update, which has undergone certain sign reversal, weight amplification, and parameter modification, the distance from the initial model will inevitably increase, at least beyond the accuracy gains it can bring on the testing dataset. 

A reliability threshold $T_\Lambda ^{iter}$ is empirically set, to construct a gap where all trainers are separated according to self-reliability. More specifically, if the self-reliability of a trainer is lower than $T_\Lambda ^{iter}$, it will be regarded as a potential malicious update, accompanying lose its aggregation qualification, i.e, its update will be removed from final aggregation process. The high-quality (i.e., high self-reliability) model updates with larger values (e.g., if it is larger than a given $T_\Lambda ^{iter}$) will be treated as pre-aggregation model waitting for final aggregation. 
\subsubsection{Attack-resistant Aggregation Algorithm}
In order to enhance the robustness of our system, we introduce an attack-resistant aggregation algorithm by adopting the repeated median estimator \cite{0Robust} and the reweighting scheme in iteratively reweighted least squares as described in \cite{2019Attack}. Specifically, we reweight each parameter by computing its residual (vertical distance) to a well-designed regression line estimated by a repeated median estimator. Assuming the $n$-th parameter of the $m$-th pre-aggregation model and the list of $n$-th parameters in all the pre-aggregation models are denoted by $y_n^m$ and $y_n$ respectively, we sort both $y_n$ and its indices $x_n$ in an ascending order. A repeated median estimator is adopted to estimate a linear regression $y = {\beta _{n0}} + {\beta _{n1}}x$, where the slope $\beta _{n1}$ and intercept $\beta _{n0}$ are estimated as
\begin{equation} 
{\beta _{n1}} = \mathop {\text{median}}\limits_i \mathop {\text{median}}\limits_{i \ne j} \frac{{y_n^j - y_n^i}}{{x_n^j - x_n^i}} \hfill. 
\end{equation} 
\begin{equation} 
{\beta _{n0}} = \mathop {\text{median}}\limits_i \mathop {\text{median}}\limits_{i \ne j} \frac{{x_n^jy_n^j - x_n^iy_n^i}}{{x_n^j - x_n^i}} \hfill. 
\end{equation}
where $i,j \in \{ 1,2,...,m\} $. The residuals of the $n$-th parameters in all the pre-aggregation models can be calculated as
\begin{equation} 
{r_n} = {y_n} - {\beta _{n0}} - {\beta _{n1}}{x_n}.
\end{equation}
Note that $r_n$ should be normalized due to different parameters has different magnitude. The normalization process can be described as
\begin{equation} 
e_n^m = \frac{{r_n^m}}{{{\tau _n}}}.
\end{equation}
where 
\begin{equation} 
{\tau _n} = \gamma  * median(\left| {{r_n}} \right|) * (1 + \frac{5}{{M - 1}}).
\end{equation}
where $\gamma$ is a constant which is set to $1.48$. Then we calculate the parameter confidence using the normalized residuals by 
\begin{equation} 
\varpi _n^m = \frac{{\sqrt {1 - {h_{mm}}} }}{{e_n^m}}\Psi (\frac{{e_n^m}}{{\sqrt {1 - {h_{mm}}} }}).
\end{equation}
where $\varpi _n^m$ is the confidence of the $n$-th parameter in $m$-th pre-aggregation model, 	$\Psi (x) = \max \left\{ { - Z,\min (Z,x)} \right\}$ with
 $Z = \upsilon \sqrt {{\raise0.7ex\hbox{$2$} \!\mathord{\left/{\vphantom {2 M}}\right.\kern-\nulldelimiterspace}\!\lower0.7ex\hbox{$M$}}} $ and $\upsilon$ is a hyperparameter, which is set to $2$. $\Psi$ here acts as a trusted interval and we can expand or shrink the interval by tuning $\upsilon$. $h_{mm}$ is the $m$-th diagonal element of matrix in $H_n$, which is expressed as ${H_n} = {x_n}{(x_n^T{x_n})^{ - 1}}x_n^T$. By involving a threshold $\varsigma$, a parameter has a confidence value lower than $\varsigma$ should be corrected as \cite{2019Attack}
 \begin{equation} 
\varpi _n^m = \varpi _n^m{\rm \mathbb{I}}(\varpi _n^m > \varsigma ).
 \end{equation}
 \begin{equation} 
y_n^m = y_n^m{\rm \mathbb{I}}(\varpi _n^m > \varsigma ) + ({\beta _{n0}} + {\beta _{n0}}x_n^m){\rm \mathbb{I}}(\varpi _n^m \leqslant \varsigma ).
 \end{equation}
 where $\mathbb{I}$ denotes a characteristic function and threshold $\varsigma$ is set to $0.1$. Finally, the weight of each pre-aggregation model can be generated by simply aggregate the parameter
 confidence of each parameter as
\begin{equation} 
{\theta ^m} = \sum\limits_{n = 1}^{N_P} {\varpi _n^m}
\label{weight}.
\end{equation}
where $N_P$ is the number of parameters. The aggregation model can be calculated as
\begin{equation} 
{w_{agg}} = \sum\limits_{m = 1}^M {\frac{{{\theta _m}}}{{\sum\nolimits_{i = 1}^M {{\theta _i}} }}} {w^m}\label{irls}.
\end{equation}

The model with corresponding quality, i.e., self-reliability and aggregate weight, of each trainer will be written into the block by consensus nodes in the form of verification transactions as shown in Figure~\ref{fig:transaction}, for following auditing. Note that once a trainer is detected as a malicious node through malicious updates remove algorithm, the aggregate weight in verification transaction will be set to $\text{-inf}$.
\begin{algorithm}[t]
	\caption{Double Aggregation Frame}
	\label{alg:DA}
	\LinesNumbered 
	\KwIn{Index of nodes $c$ and node $m$ in normal node set ${N_\ell^{iter}}$}
	\KwOut{Globe model $w^{iter}$;}
	\For{each iteration $iter$}{
		\eIf{$iter|E = 0$}
		{Each MEC server send aggregate model ${w}^{iter-1}$ to the selected trainers in charge \;
		\For{each trainer $m$}{
			Training with differential privacy to obtain differential privacy model $\tilde w_m^{iter}$\;
			Send differential privacy model $\tilde w_m^{iter}$ to associated MEC server\;
	    }
        \For{each MEC server $\ell$}{
    	Filter $\tilde w_m^{iter}$ of each received model based on self-reliability\;
    	Aggregates models of normal node set ${N_\ell^{iter}}$ as \
    	$w_\ell ^{iter} = \text{edgeagg}({\{ w_\ell ^{iter}(m)\} _{m \in N_\ell ^{iter}}})$ using Equation~(\ref{irls})\;
    	Send $w_\ell ^{iter}$ to cloud\;	
        }
        Generate ${w^{iter}} = {\text{cloudagg}}(\{ w_\ell ^{iter}\})$ as Equation~(\ref{agg})\;	
        }
		{Each trainer compute $w^{iter-1}$ by running stochastic gradient algorithm algorithm\;
		\For{each trainer $m$}{
			Training with differential privacy to obtain differential privacy model $\tilde w_m^{iter}$\;
	    }	
        }	
	}
	\textbf{Return:} ${w^{iter}}$
\end{algorithm}
\subsubsection{MEC-based Double Aggregation Frame}
Previous FL aggregation works assume one cloud server based aggregator, where millions of participated devices, e.g., vehicles in the IoV system, directly upload local updates for aggregation. However, a cloud-based aggregation approach is so sluggish that communication overhead issue may be inevitably arised with the rapid increase of terminal vehicles in IoV system. With the recent emergence of edge computing platforms, researchers have started investigating edge-based FL systems \cite{2019Client}. For edge-based FL, the proximate edge server, such as a base station, will act as the parameter server, while the clients within its communication range of the server collaborate to train a learning model, thus the latency of the computation is comparable to that of communication to the edge parameter server. Nevertheless, the finite number of vehicles each edge server can access, leading to the loss of training performance unavoidably. 

From the above comparison, we see a necessity in leveraging a cloud server to cover the massive aggregation update, while each edge server enjoys quick model updates with its local trainers. This motivates us to propose a MEC-based double aggregation frame as shown in Figure~\ref{fig:system}, where each MEC server first aggregates the filtered normal nodes surviving in malicious updates fifter based on self-reliability, to produce preliminary aggregation model, which is subsequently delivered to the cloud server to re-aggregate with the preliminary aggregation results of the remaining MEC servers. Specifically, in iteration $iter$ where $iter|E = 0$, we assume there has one cloud server, $L$ MEC servers indexed by $\ell$, with disjoint trainer sets $S_\ell ^{iter}$ and $M$ clients indexed by $m$. Each edge server firstly filters and aggregates models of normal node set ${N_\ell^{iter}}$, i.e.,
\begin{equation}
w_\ell ^{iter} = \text{edgeagg}({\{ w_\ell ^{iter}(m)\} _{m \in N_\ell ^{iter}}}).
\end{equation}
where $\text{edgeagg()}$ denotes edge aggregate function, which is expressed as Equation~(\ref{irls}). Then, the cloud server aggregates all the edge servers' models as
\begin{equation}
{w^{iter}} = {\text{cloudagg}}(\{ w_\ell ^{iter}\} ).
\end{equation}
where $\text{edgeagg()}$ denotes edge aggregate function similarly, which is expressed as Equation~(\ref{agg}). Finally, the aggregated global model will be sent to each devices via the MEC server. The details of the double aggregation frame are presented in Algorithm \ref{alg:DA}. 

Compared with cloud-based FL aggregation frame, double aggregation frame will significantly reduce the communication overhead with the cloud by aggregating massive local updates responsible by each MEC server into a preliminary model, thereby reducing the burden of backbone network. On the other hand, as more local model update that belong to different MEC servers can be accessed by the cloud server, double aggregation FL frame will outperform edge-based FL in model performance.

\subsection{Consensus: Proof of Testing Accuracy}
The consensus process is executed by all consensus nodes to competes for the qualification to write self-organized blocks to the blockchain via a consensus protocol. The node who wins the competition will be treated as a leader to broadcasts its local block to other nodes for verification. Once the verification is passed, the block is added to the permissioned blockchain which is tamper-proof. Traditional consensus protocol, e.g., PoW-based, not only brings high compution cost and communication overhead, but makes limited additional contribution to FL, thus making it less practical for the FL-based IoV system to adopt. To address this issue, we further propose a new consensus, which bases on accuracy performance of each consensus node in testing dataset, to improve the utility and efficiency of computing work in the consensus protocol. More specifically, in each iteration, once each consensus node aggregates the local updates responsible for to generate a preliminary aggregation model, we respectively predicts the testing dataset using aggregation model of all consensus nodes, and finally the consensus node with the highest testing dataset consensus accuracy will be selected as the leader. Generally, the consensus accuracy ${\bar \alpha _\ell }$ of each consensus node $\ell$ is denoted by the fraction of correctly classified samples. The leader, which gathers all the transactions it received to form a local block $B_{leader}$, is responsible for driving the consensus process among consensus nodes. $B_{leader}$ is broadcasted to all consensus nodes of blockchain for approval. If the block containing all transactions is approved by every consensus node, the block data signed with leader's signature will be stored in the blockchain, which are tamperproof. Similar to other blockchain systems, the leader will get a certain block reward, which is denoted as ${\rho _b}$.

\subsection{Aggregation Weight based Incentive Mechanism}
In the current federated learning paradigm, most existing works assume optimistically that all the vehicles in IoV system will participate in FL tasks unconditionally when were invited. However, it is not practical in the real world due to resource costs emerged during model training. Thus, it is vital to bring in some incentive mechanism that maps the contributed resources into appropriate rewards. 
In this paper, we address the problem of treating FL participants fairly based on their contributions, which can be quantified by aggregation weight as shown in equation~(\ref{weight}), to build a healthy FL incentive mechanism. 

Specifically, in iteration $iter$ where $iter|E = 0$, we assume that the set of selected trainers in consensus node $\ell$ is denoted by $T_\ell^{iter}$ and define the contribution of each participant $m$ as follows:
\begin{equation}
C_m^{iter} = \left\{ \begin{gathered}
0\quad\;\;\;\;{\text{       if  }}m \notin {T^{iter}}{\text{ or  }}m \notin {\mathbb{N}_\ell^{iter}} \hfill \\
\theta_m^{iter}{\;\;\text{  if  }}m \in {\mathbb{N}_\ell^{iter}}{\text{ }} \hfill \\ 
\end{gathered}  \right.
\end{equation}
where ${\mathbb{N}_\ell^{iter}}$ represents the set of normal trainers in consensus node $\ell$. Note that the contribution of each participant stored transparently in the blockchain, which means that every blockchain node can audit the authenticity of contributions. Thence the total contribution for device $m$ in a complete FL task can be calculated as:
\begin{equation}
{C_m} = \sum\limits_{iter = 1}^K {C_m^{iter}}. 
\end{equation}
where $K$ denotes the total aggregation times.

We distribute the available payoffs  ${\rho _{ava}}$ to each consensus node $\ell$ in proportion to the consensus accuracy ${\bar \alpha _\ell }$, i.e., 
\begin{equation}
{\rho _\ell } = {\rho _{ava}} * \frac{{{{\bar \alpha }_\ell }}}{{\sum\nolimits_{i = 1}^\ell  {{{\bar \alpha }_i}} }}.
\end{equation}
where ${\rho _\ell }$ denotes the distribute payoffs of consensus node $\ell$. In this paper, we only consider two payment output, i.e., block reward and participant payoff, thence, ${\rho _{ava}} = {\rho _{total}} - K * {\rho _b}$. Intuitively, devices shall receive payoffs that is proportional to their contributions. Therefore, the total payoffs of a device $m$ within consensus node gainning from FL task  $\ell$ should be expressed as:
\begin{equation}
{\rho _m} = {\rho _\ell } * \frac{{{C_m}}}{{\sum\limits_{m = 1}^M {{C_m}} }}.
\end{equation}
\section{Analysis and Evaluation}
This section revisits our design goals of BMFL presented in section IV and gives performance analysis accordingly. Then we implement the prototype of BMFL
and evaluate its performance with benchmark, open real-world datasets for data categorization.
\subsection{Performance Analysis}
\subsubsection{Achieving Differential Privacy}
According to standard argument \cite{2013The}, if we choose $\sigma  \geqslant \frac{{\sqrt {2\ln ({{1.25} \mathord{\left/{\vphantom {{1.25} \delta }} \right.\kern-\nulldelimiterspace} \delta })}}}{\varepsilon }$ in Algorithm \ref{alg:DP}, each step of local training will follows the requirement of differential privacy, i.e., Equation~(\ref{DP}) and the final results will satisfy $(\varepsilon,\delta)$-differential privacy. Notice that the privacy budget is only cost in Step $14$, in which gaussian noise is added to clipped gradient. Therefore, Algorithm \ref{alg:DP} satisfies $(\varepsilon,\delta)$-differential privacy.
\subsubsection{Reducing Communication Overhead}
Compared with cloud-based FL aggregation frame, our double aggregation frame will significantly reduce the costly communication with the cloud by aggregating massive local model updates responsible by the MEC server into a preliminary model, thereby, reducing the burden of backbone network. Regarding the communication issue, our proposed methods reduce the up-stream communication cost with the cloud from $d * K$ floats (model size times number of trainer) to $d * \ell$ floats (model size times number of MEC server) in one aggregation round.
\subsubsection{Enhancing Robustness}
Although residual based reweighting aggregation algorithm, which calculates the confidence of updates based on the residual to a joint repeated median of all models, can achieve good performance in non-IID situations. However, due to the uncertainty in the distribution of malicious trainers, e.g., the proportion of malicious trainers in the trainer-set responsible for of a certain consensus node is likely to exceed 51$\%$, the residual based reweighting aggregation algorithm will estimate a wrong median linear regression standard, leading to failure of robust aggregation. By contrast, our self-reliability based filter does not depend on the model parameters of other training nodes, but is only related to the quality of its own model, thus naturally avoiding the influence of 51$\%$ malicious models proportion. Simply put, our method is equivalent to adding another layer of protection to reweighting aggregation algorithm, thereby achieving a more robust enhancement.
\subsubsection{Fair Incentive Mechanism}
Aggregation weight based incentive mechanism determines that the upload model with poor quality which tends to have a lower aggregation weight, will get less profit, and vice versa. The detected malicious nodes will not distribute any profits, thereby suppressing malicious attacks. Moreover, we distribute the available profits ${\rho _{ava}}$ to each consensus node $\ell$ in proportion to the consensus accuracy ${\bar \alpha _\ell }$, so as to achieve a more fair distribution of profits. 
\subsection{Evaluation Setup}
We consider a IoV system with $50$ devices, 5 MEC servers and a cloud server, assuming each edge server authorizes the same number of clients. A random generator is used to randomly assign 10 devices to each MEC server. We ignore the
possible heterogeneous communication conditions and computing resources for different clients. Without loss of generality, we conduct evaluations where we compare our approach with other algorithms, including FedAvg  \cite{2016Communication}, and residual based reweighting aggregation algorithm \cite{2019Attack} on a real-world dataset MNIST, which contains 70,000 real-world handwritten images with digits from $0$ to $9$. The comprehensive federated setting is divided into two ways in partitioning the MNIST data over devices, i.e, 1) complete non-IID way, where we distribute the data among 1,000 devices such that each device has samples of only two digits and the number of samples per device follows a power law, 2) quantity non-IID way, where we adopt a Dirichlet distribution with a hyperparameter 0.9 to generate non-IID data distribution for totally 50 devices.

We evaluate different methods by learning a global model, i.e., a two-layer convolutional neural network with 21840 trainable parameters. With this simple CNN model, our goal is to evaluate different algorithms for defending federated learning in the presence of malicious attacks. In this paper, We consider the following attack types:
\begin{itemize}
	\item \textbf{\textit{Sign-flipping attack.}} Sign-flipping attack \cite{li2019rsa} is an untargeted attack, where the malicious clients flip the signs of their local model updates by multiplying with a negative constant $\gamma $ and sends to model aggregators. In this paper, we set $\gamma = -10$.
	Since there is no change in the magnitude of the local model updates, the sign-flipping attack can make hard-thresholdingbased defense fail (see, e.g., \cite{sun2019can}). In this experience, we partition the MNIST data over devices in a complete non-IID way.
	\textbf{\item \textit{Same-value attack.}}
	Same-value attack \cite{peng2020byzantine} is also an untargeted attack where the malicious clients send the local model update as
	$\tilde w_m^{iter} = c*I$ to model aggregators. Here $I$ indicates an all-in-one matrix with the same dimension as the original local model update and $c$  is a constant which we set as $100$. In this experience, we partition the MNIST data over devices in a complete non-IID way.
	\textbf{\item \textit{Backdoor attack.}}
	Backdoor attack is targeted attack, a.k.a. model poisoning attack \cite{bagdasaryan2020backdoor}, aiming to change an ML model’s behaviours on a minority of data items while maintaining the primary model performance across the whole testing dataset. Similar with \cite{2019Attack}, a quantity non-IID way is adopted to partition the MNIST data over devices.
\end{itemize}
In each round of federated learning, each participant is supposed to train the local model for 5 epochs, but the attackers can train for arbitrary epochs. 
The federated setting of our experimental scenario is as
\begin{itemize}
 	\item \textbf{\textit{Learning rate of normal trainers $\eta = 0.01$}}.
 	\item \textbf{\textit{Learning rate of attackers $\eta_a = 0.1$}}.
 	\textbf{\item \textit{Local batchsize $B = 64$}}.
 	\textbf{\item \textit{Proportion of trainers $V = 1$}}.
 	\textbf{\item \textit{Number of normal trainers local epochs $E = 5$}}.
 	\textbf{\item \textit{Number of attackers local epochs $E_a = 10$}}.
 	\textbf{\item \textit{SGD momentum $\beta = 0.9$}}.
 	\textbf{\item \textit{DP-based FL respectively fixes $C = 1,\sigma = 2, \delta = 0.00001$}}.
\end{itemize}
We implement our simulation experiments on the pytorch platform where a computer with the processor of Intel Core i5-8400 CPU@3.20GHz and RAM 8.0GB are installed.
\subsection{Evaluation Result}
\begin{figure}[!t]
	\centerline{\includegraphics[width=1\linewidth]{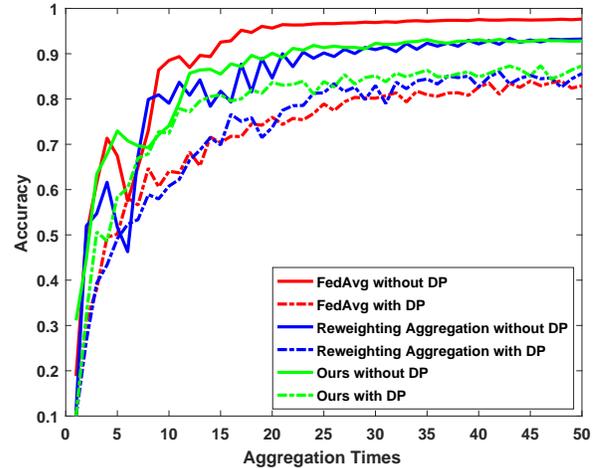}}
	\caption{Comparison in Testing Accuracy with different methods under differential privacy and without differential privacy.}
	\label{fig:DP}
\end{figure}
\begin{figure*}[t]
	\centering	
	\subfigure[Testing accuracy under $20\%$ sign-flipping attackers.]{
		\includegraphics[width=0.31\textwidth]{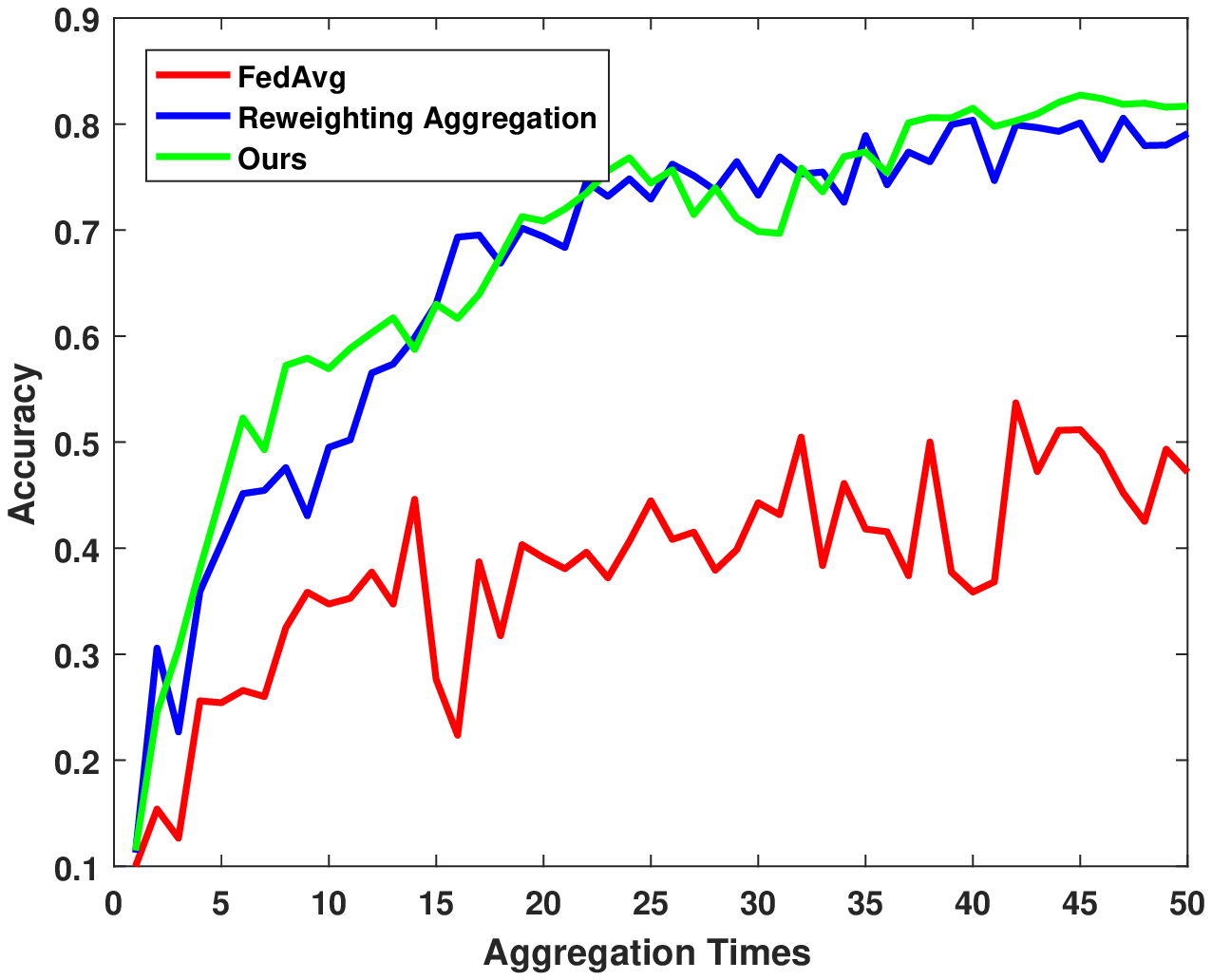}}
	~	
	\subfigure[Testing accuracy under $40\%$ sign-flipping attackers.]{
		\includegraphics[width=0.31\textwidth]{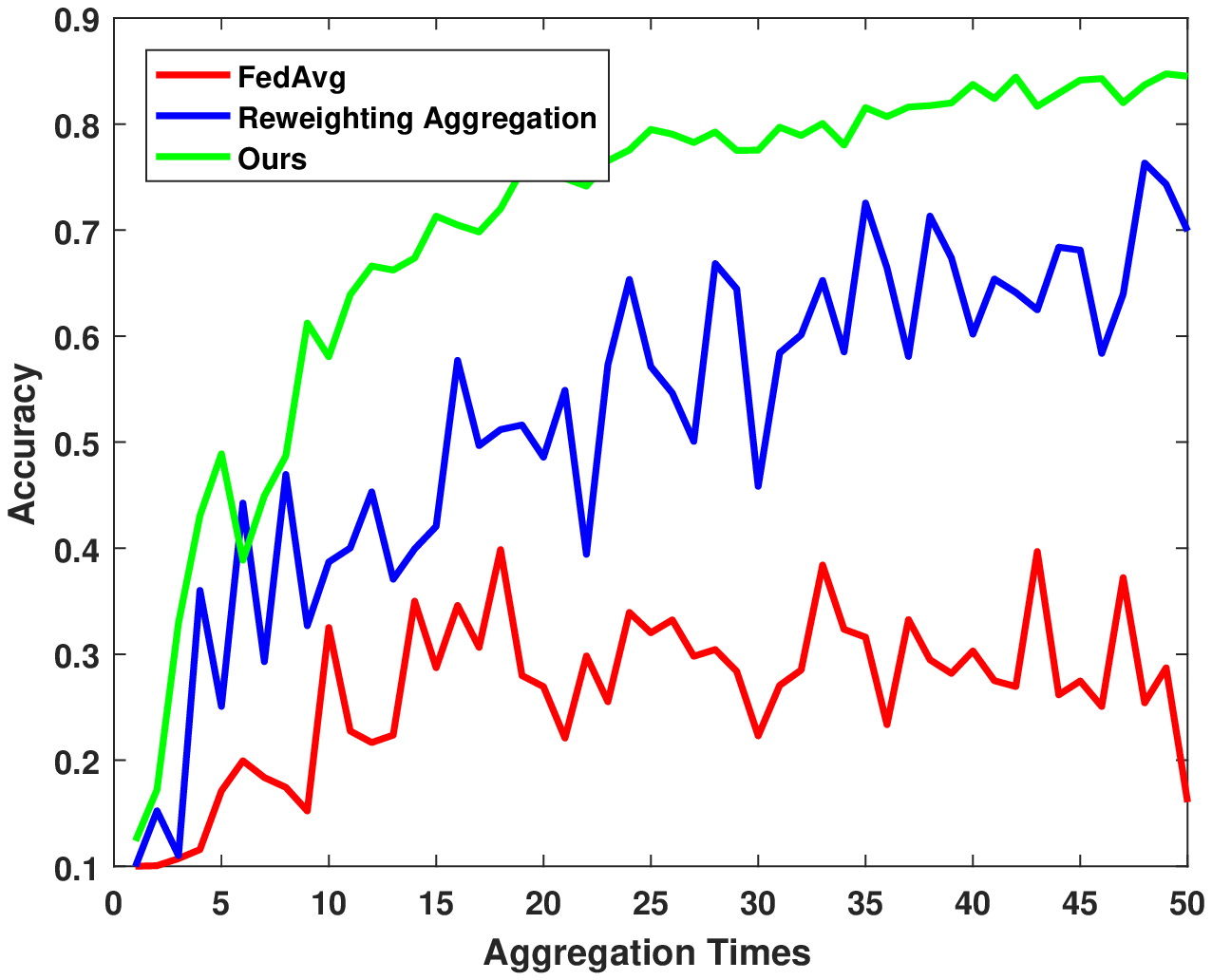}}
	~
	\subfigure[Testing accuracy under $60\%$ sign-flipping attackers.]{
		\includegraphics[width=0.31\textwidth]{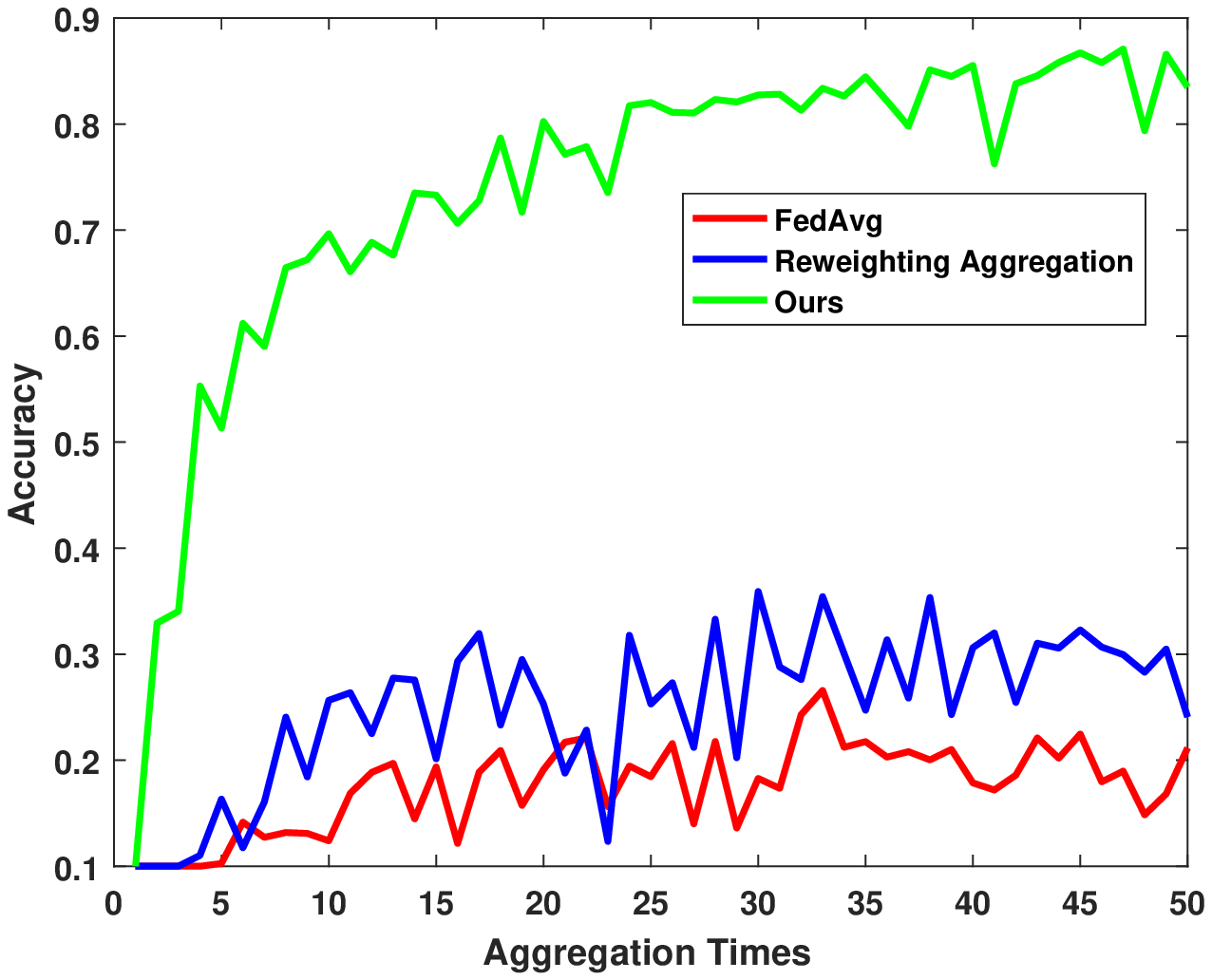}}
	~
	\subfigure[Testing accuracy under $20\%$ same-value attackers.]{
		\includegraphics[width=0.31\textwidth]{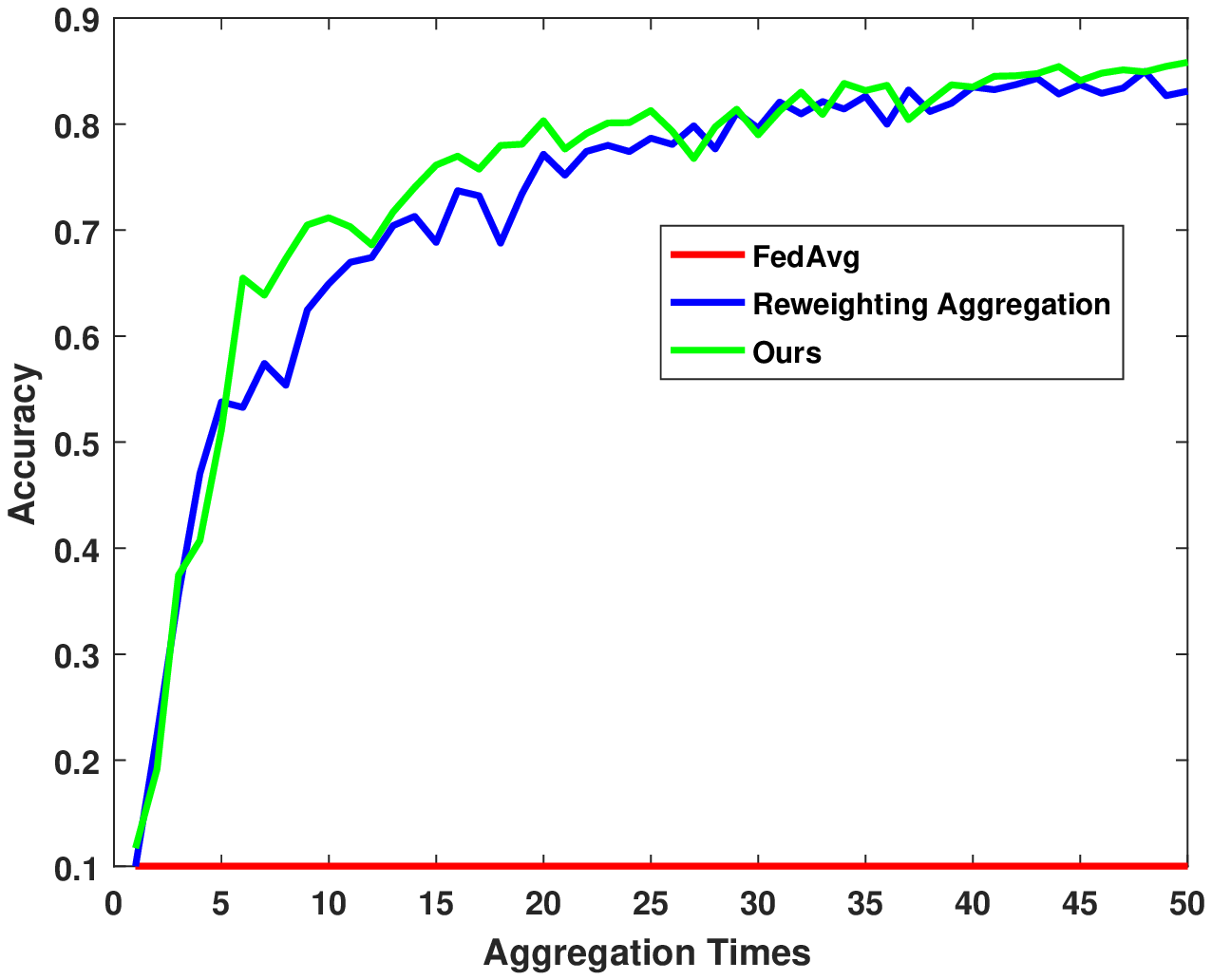}}
	~	
	\subfigure[Testing accuracy under $40\%$ same-value attackers.]{
		\includegraphics[width=0.31\textwidth]{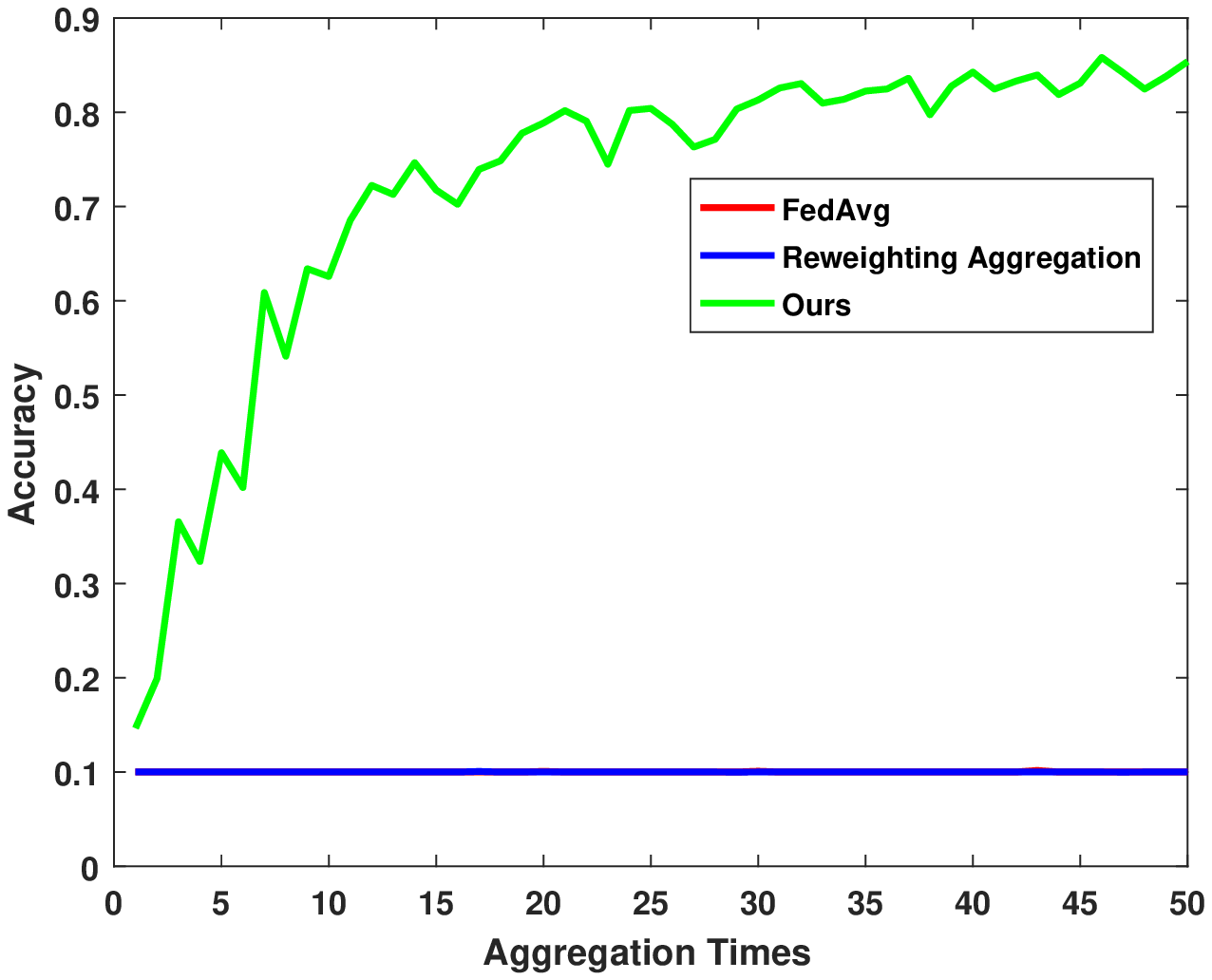}}
	~
	\subfigure[Testing accuracy under $60\%$ same-value attackers.]{
		\includegraphics[width=0.31\textwidth]{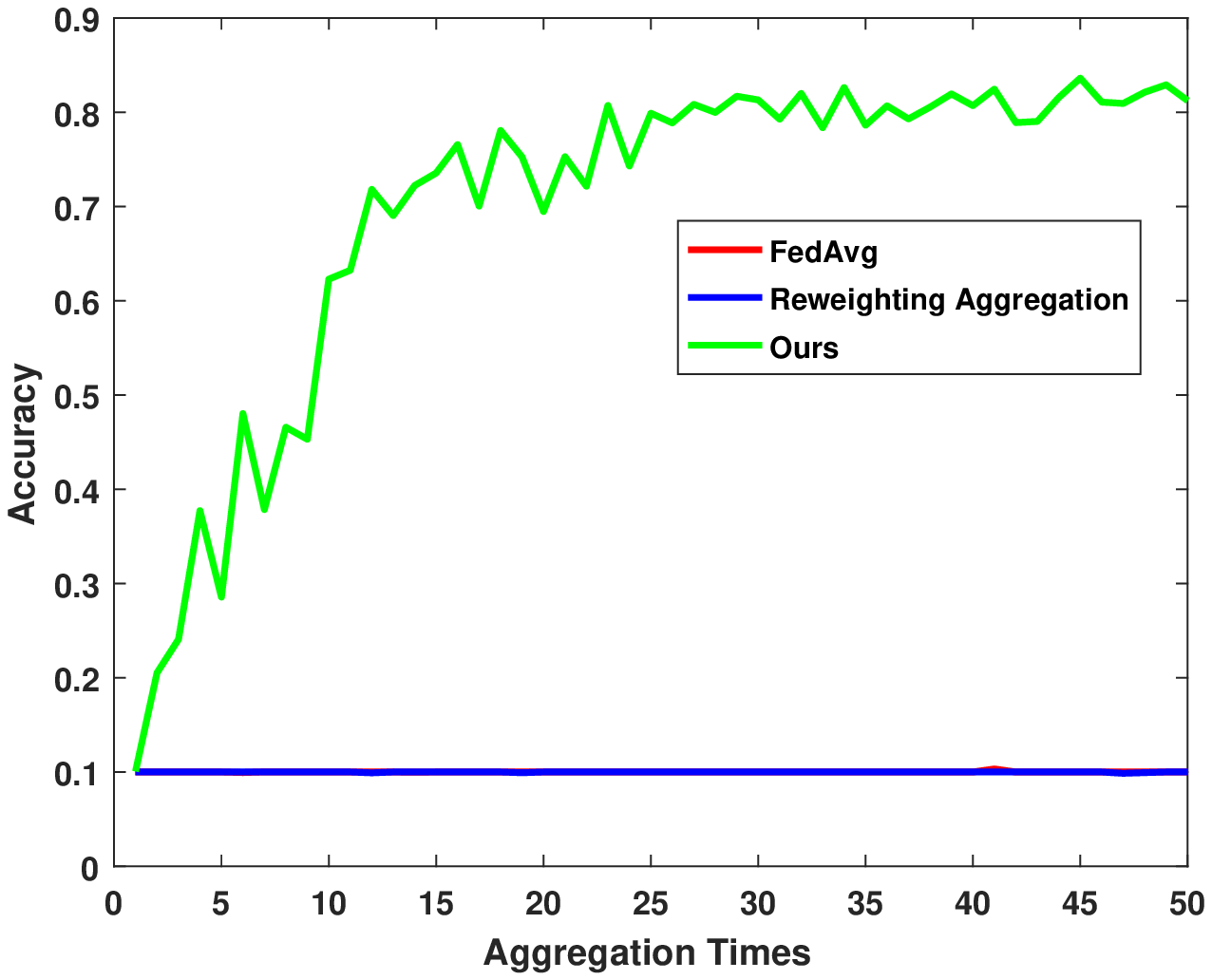}}
	~
	\caption{Sign-flipping and same-value attack on FedAvg, reweighting aggregation algorithm, and our proposed method with the MNIST dataset depending on the fraction of malicious attacks.}
	\label{fig:sign-flipping}
\end{figure*}
\begin{figure*}[t]
	\centering	
	\subfigure[Attack success rate under $40\%$ backdoor attackers.]{
		\includegraphics[width=0.31\textwidth]{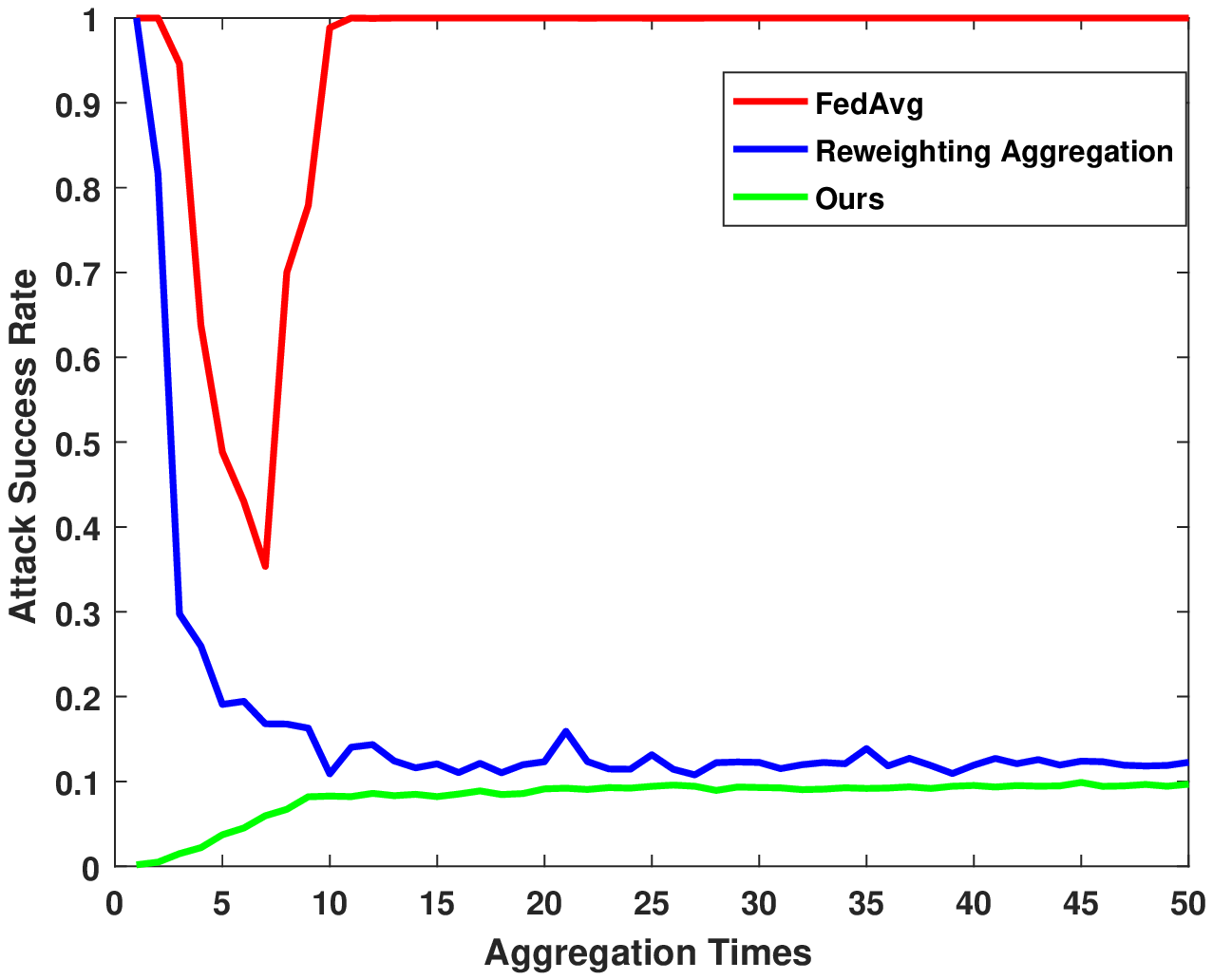}}
	~
	\subfigure[Attack success rate under $50\%$ backdoor attackers.]{
		\includegraphics[width=0.31\textwidth]{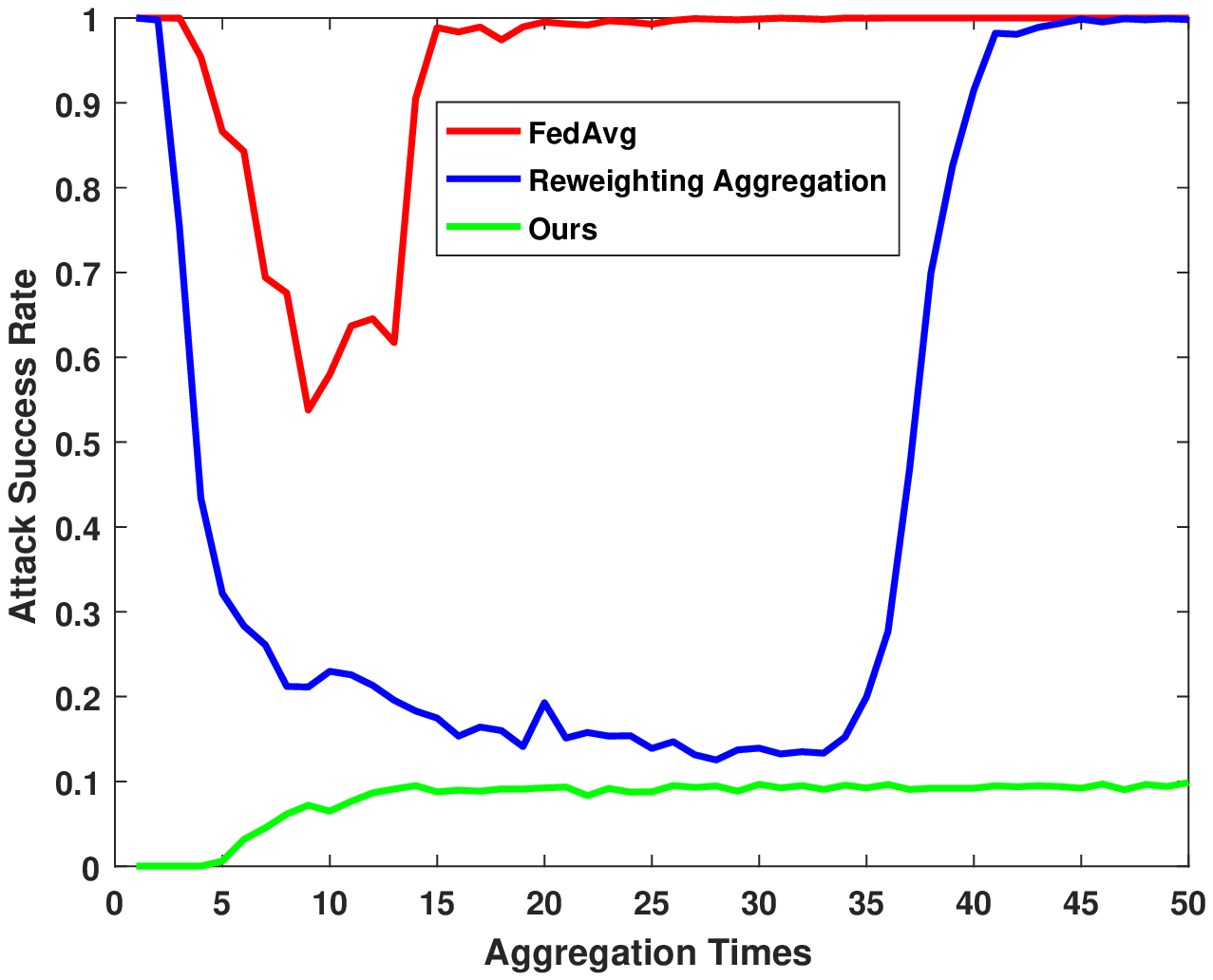}}
	~
	\subfigure[Attack success rate under $60\%$ backdoor attackers.]{
		\includegraphics[width=0.31\textwidth]{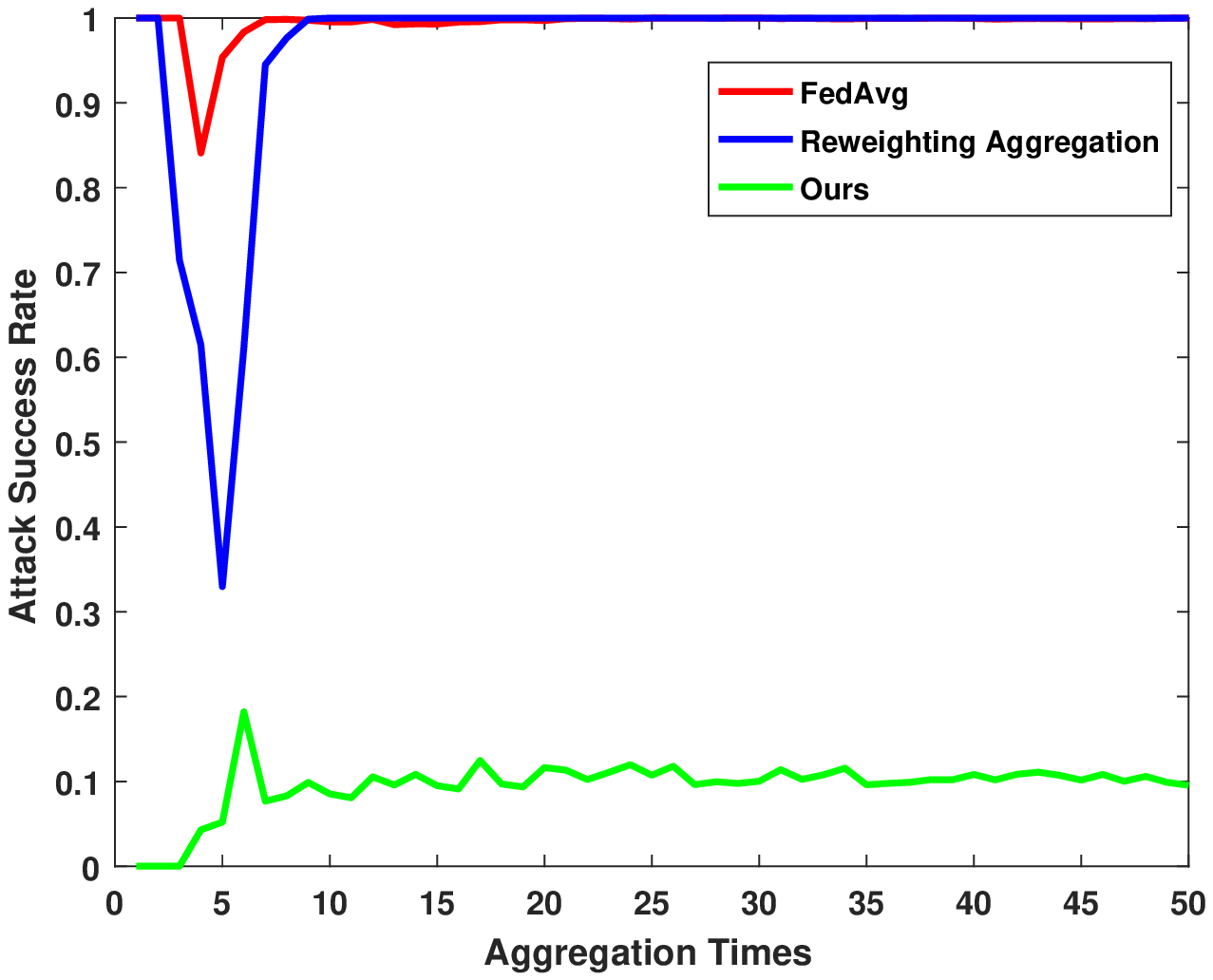}}
	~	
	\caption{Backdoor attack on FedAvg, reweighting aggregation algorithm, and our proposed method with the MNIST dataset depending on the fraction of malicious attacks.}
	\label{fig:backdoor}
\end{figure*}
\subsubsection{Performance under Differential Privacy}
We draw the testing accuracy as a function of aggregation times with different methods under differential privacy and without differential privacy, firstly. The comparative results are shown in Figure~\ref{fig:DP}. As we can see, testing accuracy are reducing as the differential privacy is adopted. The reason is the differential privacy-based FL schemes preserve privacy by adding noises into original gradients. In order to improve the level of privacy protection, it is necessary to add noises with a variance $\sigma$, thus satisfying $(\varepsilon,\delta)$-differential privacy. However, noises with a large variance have a more significant negative impact on testing accuracy. 
\subsubsection{Performance under Sign-flipping Attack and Same-value Attack}
We evaluate the overall classification performance of different methods under sign-flipping attack and same-value attack within differential privacy. Experimental results are shown in Figure~\ref{fig:sign-flipping}. The results show that our approach outperforms FedAvg and reweighting aggregation algorithm in defending the sign-flipping attack and same-value Attack when the proportion of malicious attackers is gradually increasing. More specifically, due to involve model information of all malicious attackers in model aggregation, FedAvg fails to converge and has relatively low accuracy under both sign-flipping and same-value attack. For instance, when sign-flipping attack exists, the achievable accuracy of FedAvg is less than $50\%$ even if $20\%$ of trainer are malicious, and roughly $40\%$ and $30\%$ under $40\%$ attackers and $60\%$ attackers respectively. The performance of our method is similar to reweighting aggregation algorithm in the presence of $20\%$ attackers. However, as the proportion of attackers increases, i.e., $40\%$ and $60\%$, reweighting aggregation algorithm gradually faces the issue of convergence failure under both sign-flipping and same-value attack. Moreover, the achievable accuracy of reweighting aggregation algorithm under same-value attack is only $10\%$ compared with sign-flipping attack. The reason can be summarized as, reweighting aggregation algorithm which calculates the confidence of updates based on the residual to a joint repeated median of all gradients, thus resulting  in the linear regression line will be determined by the malicious models when the proportion of malicious attackers exceeds a certain ratio. In this case, most of the malicious updates will be aggregated, which will cause the global model to face convergence failure. Notice that same-value attack greatly amplifies the model parameters, leading the global model deviates further from the normal update trajectory. Thence, same-value attack may cause more significant damage on the model performance than sign-flipping attack.
\subsubsection{Performance under Backdoor Attack}
For pixel-pattern backdoor attacks in federated learning, attackers manipulate their local models so that the learned global model predicts some backdoor target label for any image embedded with certain patterns. The global model can behave normally for clean data. We choose ``2'' in MNIST as the backdoor target labels. The training data is mixed with manipulated data and clean data to fit both the backdoor task and the main task. Morever, an attacker poisons its local model and submits the malicious update in every round. 
Figure~\ref{fig:backdoor} summarizes the results in attack success rate of backdoor attacks over time under different algorithms. Obviously, our method achieves the lowest in terms of attack success rate on MNIST under both backdoor attack scenarios. Intuitively, backdoor attacks can easily succeed under FedAvg due to absorb all of local updates with backdoor attacks and another baseline, i.e., reweighting aggregation algorithm, only works when there are a small number of attackers. When the proportion of attackers increases, reweighting aggregation algorithm can only slow down the process but still reach high attack success
rate into over $100\%$ within 50 aggregation times. By contrast, our algorithm effectively defends the backdoor attack and remains stable with $9.54\%$ attack success rate when being attacked continuously for 50 aggregation times and the proportion of attackers is increasing.
\section{Conclusion}
In this paper, we addressed data sharing issues related to ensuring secure, effective robust IoV system, by integrating blockchain, differential privacy, and MEC technology to federated learning framework. Specifically, local differential privacy techniques is introduced to prevent privacy concerns while the auditability of update models are ensured by  based on self-reliability filter. In addition, a blockchain architecture is employed to build a federated learning participant management system and ensure the immutability of uploaded model, whose quality quantified as aggregate weight is used as a criterion for the distribution of federal mission profits. Simultaneously, an double aggregation frame based on mobile edge computing is proposed to effectively reduce cloud communication overhead and ensure the quality of model training. The results of analyses and experiments demonstrate that the proposed data sharing scheme achieves good accuracy, high robustness, and enhanced security.
\bibliographystyle{IEEEtran}      
\bibliography{ref}

\begin{thebibliography}{10}
\providecommand{\url}[1]{#1}
\csname url@samestyle\endcsname
\providecommand{\newblock}{\relax}
\providecommand{\bibinfo}[2]{#2}
\providecommand{\BIBentrySTDinterwordspacing}{\spaceskip=0pt\relax}
\providecommand{\BIBentryALTinterwordstretchfactor}{4}
\providecommand{\BIBentryALTinterwordspacing}{\spaceskip=\fontdimen2\font plus
\BIBentryALTinterwordstretchfactor\fontdimen3\font minus
  \fontdimen4\font\relax}
\providecommand{\BIBforeignlanguage}[2]{{%
\expandafter\ifx\csname l@#1\endcsname\relax
\typeout{** WARNING: IEEEtran.bst: No hyphenation pattern has been}%
\typeout{** loaded for the language `#1'. Using the pattern for}%
\typeout{** the default language instead.}%
\else
\language=\csname l@#1\endcsname
\fi
#2}}
\providecommand{\BIBdecl}{\relax}
\BIBdecl

\bibitem{feng2019toward}
D.~Feng, C.~She, K.~Ying, L.~Lai, Z.~Hou, T.~Q. Quek, Y.~Li, and B.~Vucetic,
  ``Toward ultrareliable low-latency communications: Typical scenarios,
  possible solutions, and open issues,'' \emph{IEEE Vehicular Technology
  Magazine}, vol.~14, no.~2, pp. 94--102, 2019.

\bibitem{sun2019blockchain}
Y.~Sun, L.~Zhang, G.~Feng, B.~Yang, B.~Cao, and M.~A. Imran,
  ``Blockchain-enabled wireless internet of things: Performance analysis and
  optimal communication node deployment,'' \emph{IEEE Internet of Things
  Journal}, vol.~6, no.~3, pp. 5791--5802, 2019.

\bibitem{2017A}
Y.~Mao, C.~You, J.~Zhang, K.~Huang, and K.~B. Letaief, ``A survey on mobile
  edge computing: The communication perspective,'' \emph{IEEE Communications
  Surveys and Tutorials}, vol.~PP, no.~99, pp. 1--1, 2017.

\bibitem{2018Learning}
H.~Li, K.~Ota, and M.~Dong, ``Learning iot in edge: Deep learning for the
  internet of things with edge computing,'' \emph{IEEE Network}, vol.~32,
  no.~1, pp. 96--101, 2018.

\bibitem{2016Federated}
J.~Konen, H.~B. Mcmahan, D.~Ramage, and P.~Richtárik, ``Federated
  optimization: Distributed machine learning for on-device intelligence,''
  \emph{arXiv preprint arXiv:1610.02527}, 2016.

\bibitem{9079513}
S.~R. {Pokhrel} and J.~{Choi}, ``Federated learning with blockchain for
  autonomous vehicles: Analysis and design challenges,'' \emph{IEEE
  Transactions on Communications}, vol.~68, no.~8, pp. 4734--4746, 2020.

\bibitem{bagdasaryan2020backdoor}
E.~Bagdasaryan, A.~Veit, Y.~Hua, D.~Estrin, and V.~Shmatikov, ``How to backdoor
  federated learning,'' in \emph{International Conference on Artificial
  Intelligence and Statistics}, 2020, pp. 2938--2948.

\bibitem{2019Client}
T.~Nishio and R.~Yonetani, ``Client selection for federated learning with
  heterogeneous resources in mobile edge,'' in \emph{ICC 2019 - 2019 IEEE
  International Conference on Communications (ICC)}, 2019.

\bibitem{2016Communication}
H.~B. Mcmahan, E.~Moore, D.~Ramage, S.~Hampson, and B.~A.~y. Arcas,
  ``Communication-efficient learning of deep networks from decentralized
  data,'' 2016.

\bibitem{nakamoto2019bitcoin}
S.~Nakamoto, ``Bitcoin: A peer-to-peer electronic cash system,'' Manubot, Tech.
  Rep., 2019.

\bibitem{wood2014ethereum}
G.~Wood \emph{et~al.}, ``Ethereum: A secure decentralised generalised
  transaction ledger,'' \emph{Ethereum project yellow paper}, vol. 151, no.
  2014, pp. 1--32, 2014.

\bibitem{9098045}
L.~{Lyu}, J.~{Yu}, K.~{Nandakumar}, Y.~{Li}, X.~{Ma}, J.~{Jin}, H.~{Yu}, and
  K.~S. {Ng}, ``Towards fair and privacy-preserving federated deep models,''
  \emph{IEEE Transactions on Parallel and Distributed Systems}, vol.~31,
  no.~11, pp. 2524--2541, 2020.

\bibitem{androulaki2018hyperledger}
E.~Androulaki, A.~Barger, V.~Bortnikov, C.~Cachin, K.~Christidis, A.~De~Caro,
  D.~Enyeart, C.~Ferris, G.~Laventman, Y.~Manevich \emph{et~al.}, ``Hyperledger
  fabric: a distributed operating system for permissioned blockchains,'' in
  \emph{Proceedings of the thirteenth EuroSys conference}, 2018, pp. 1--15.

\bibitem{larimer2013transactions}
D.~Larimer, ``Transactions as proof-of-stake,'' \emph{Nov-2013}, 2013.

\bibitem{castro1999practical}
M.~Castro, B.~Liskov \emph{et~al.}, ``Practical byzantine fault tolerance,'' in
  \emph{OSDI}, vol.~99, no. 1999, 1999, pp. 173--186.

\bibitem{FERNANDO201384}
``Mobile cloud computing: A survey,'' \emph{Future Generation Computer
  Systems}, vol.~29, no.~1, pp. 84 -- 106, 2013, including Special section:
  AIRCC-NetCoM 2009 and Special section: Clouds and Service-Oriented
  Architectures.

\bibitem{8016573}
Y.~{Mao}, C.~{You}, J.~{Zhang}, K.~{Huang}, and K.~B. {Letaief}, ``A survey on
  mobile edge computing: The communication perspective,'' \emph{IEEE
  Communications Surveys and Tutorials}, vol.~19, no.~4, pp. 2322--2358, 2017.

\bibitem{2018Inference}
L.~Melis, C.~Song, E.~De~Cristofaro, and V.~Shmatikov, ``Inference attacks
  against collaborative learning,'' 2018.

\bibitem{hitaj2017deep}
B.~Hitaj, G.~Ateniese, and F.~Perez-Cruz, ``Deep models under the gan:
  information leakage from collaborative deep learning,'' in \emph{Proceedings
  of the 2017 ACM SIGSAC Conference on Computer and Communications Security},
  2017, pp. 603--618.

\bibitem{8668426}
Y.~{Sun}, L.~{Zhang}, G.~{Feng}, B.~{Yang}, B.~{Cao}, and M.~A. {Imran},
  ``Blockchain-enabled wireless internet of things: Performance analysis and
  optimal communication node deployment,'' \emph{IEEE Internet of Things
  Journal}, vol.~6, no.~3, pp. 5791--5802, 2019.

\bibitem{2016Hawk}
A.~Kosba, A.~Miller, E.~Shi, Z.~Wen, and C.~Papamanthou, ``Hawk: The blockchain
  model of cryptography and privacy-preserving smart contracts,'' in
  \emph{Security \& Privacy}, 2016.

\bibitem{2019Blockchain}
Y.~Lu, X.~Huang, Y.~Dai, S.~Maharjan, and Y.~Zhang, ``Blockchain and federated
  learning for privacy-preserved data sharing in industrial iot,'' \emph{IEEE
  Transactions on Industrial Informatics}, vol.~PP, no.~99, pp. 1--1, 2019.

\bibitem{liu2020secure}
Y.~Liu, J.~Peng, J.~Kang, A.~M. Iliyasu, D.~Niyato, and A.~A.~A. El-Latif, ``A
  secure federated learning framework for 5g networks,'' \emph{arXiv preprint
  arXiv:2005.05752}, 2020.

\bibitem{dwork2014algorithmic}
C.~Dwork, A.~Roth \emph{et~al.}, ``The algorithmic foundations of differential
  privacy.'' \emph{Foundations and Trends in Theoretical Computer Science},
  vol.~9, no. 3-4, pp. 211--407, 2014.

\bibitem{abadi2016deep}
M.~Abadi, A.~Chu, I.~Goodfellow, H.~B. McMahan, I.~Mironov, K.~Talwar, and
  L.~Zhang, ``Deep learning with differential privacy,'' in \emph{Proceedings
  of the 2016 ACM SIGSAC Conference on Computer and Communications Security},
  2016, pp. 308--318.

\bibitem{zhuang2019false}
P.~Zhuang, R.~Deng, and H.~Liang, ``False data injection attacks against state
  estimation in multiphase and unbalanced smart distribution systems,''
  \emph{IEEE Transactions on Smart Grid}, vol.~10, no.~6, pp. 6000--6013, 2019.

\bibitem{zhu2018blockchain}
X.~Zhu, H.~Li, and Y.~Yu, ``Blockchain-based privacy preserving deep
  learning,'' in \emph{International Conference on Information Security and
  Cryptology}.\hskip 1em plus 0.5em minus 0.4em\relax Springer, 2018, pp.
  370--383.

\bibitem{li2019rsa}
L.~Li, W.~Xu, T.~Chen, G.~B. Giannakis, and Q.~Ling, ``Rsa: Byzantine-robust
  stochastic aggregation methods for distributed learning from heterogeneous
  datasets,'' in \emph{Proceedings of the AAAI Conference on Artificial
  Intelligence}, vol.~33, 2019, pp. 1544--1551.

\bibitem{2019Attack}
S.~Fu, C.~Xie, B.~Li, and Q.~Chen, ``Attack-resistant federated learning with
  residual-based reweighting,'' 2019.

\bibitem{0Robust}
A.~F. Siegel, ``Robust regression using repeated medians,'' \emph{Biometrika},
  no.~1, pp. 242--244.

\bibitem{2013The}
C.~Dwork and A.~Roth, ``The algorithmic foundations of differential privacy,''
  \emph{Foundations and Trends in Theoretical Computer ence}, vol.~9, no. 3-4,
  pp. 1--27, 2013.

\bibitem{sun2019can}
Z.~Sun, P.~Kairouz, A.~T. Suresh, and H.~B. McMahan, ``Can you really backdoor
  federated learning?'' \emph{arXiv preprint arXiv:1911.07963}, 2019.

\bibitem{peng2020byzantine}
J.~Peng and Q.~Ling, ``Byzantine-robust decentralized stochastic optimization
  over static and time-varying networks,'' \emph{arXiv preprint
  arXiv:2005.06276}, 2020.

\end{thebibliography}

\end{document}